\documentclass[aps,prx,reprint]{revtex4-2}

\usepackage{amsmath,amssymb,amsfonts}
\usepackage{graphicx}
\usepackage{epsfig}
\usepackage{color}
\usepackage{esint}
\DeclareUnicodeCharacter{3B1}{\ensuremath{\alpha}}
\DeclareUnicodeCharacter{3B2}{\ensuremath{\beta}}

\usepackage{xcolor}  

\begin{document}
	
	
	\title{Active Hydrodynamic Theory of Euchromatin and Heterochromatin}
	
	
	\author{S.\ Alex Rautu$^1$}
	
	\author{Alexandra Zidovska$^2$}
	
	\author{David Saintillan$^3$}
	
	\author{Michael J.\ Shelley$^{1,4}$}
	\affiliation{
		$^1$Center for Computational Biology, Flatiron Institute, New York, NY 10010, USA\\
		$^2$Center for Soft Matter Research, Department of Physics, New York University, New York, NY 10003, USA\\
		$^3$Department of Mechanical and Aerospace Engineering, University of California, San Diego, 9500 Gilman Drive, La Jolla, California 92093, USA\\
		$^4$Courant Institute, New York University, New York, NY 10012, USA
	}
	
	\begin{abstract}
		The genome contains genetic information essential for cell's life. The genome's spatial organization inside  the cell nucleus is critical for its proper function including gene regulation. The two major genomic compartments -- euchromatin and heterochromatin -- contain largely transcriptionally active and silenced genes, respectively, and exhibit distinct dynamics. In this work, we present a hydrodynamic framework that describes the large-scale behavior of euchromatin and heterochromatin, and accounts for the interplay of mechanical forces, active processes, and nuclear confinement. Our model shows contractile stresses from cross-linking proteins lead to the formation of heterochromatin droplets via mechanically driven phase separation. These droplets grow, coalesce, and in nuclear confinement, wet the boundary. Active processes, such as gene transcription in euchromatin, introduce non-equilibrium fluctuations that drive long-range, coherent motions of chromatin as well as the nucleoplasm,  and thus alter the genome's spatial organization. These fluctuations also indirectly deform heterochromatin droplets, by continuously changing their shape. Taken together, our findings reveal how active forces, mechanical stresses and hydrodynamic flows contribute to the genome's organization at large scales and provide a physical framework for understanding chromatin organization and dynamics in live cells.
	\end{abstract}
	
	
	
	
	\maketitle

	\section{Introduction}
	
	The spatial organization of the genome inside the cell nucleus is critical for the genome's proper function~\cite{Alberts2008, Cremer2001, Misteli2007, Bickmore2013, Gibcus2013, Pombo2015, Mirny2022, Bonev2016}. The genome is present in the cell in the form of chromatin, a complex of DNA and histone proteins~\cite{Alberts2008}. Chromatin exists in two major structural and functional states: euchromatin, which is loosely packed and transcriptionally active, and heterochromatin, which is densely packed and transcriptionally silent \cite{Bonev2016, Belmont2017, Solovei2016}. These two chromatin states are spatially segregated, with euchromatin generally located in the nuclear interior and heterochromatin near the nuclear periphery and nucleolar surface \cite{Bizhanova2021, Volk2021, Solovei2020b}. This compartmentalization is a hallmark of nuclear organization across a wide range of cell types and plays a key role in gene regulation, affecting processes such as gene expression, genome replication and DNA damage repair \cite{Solovei2016, Gibcus2013, Lopes2024, Liu2018}.
	
	The genome's organization arises from a multifaceted interplay of biochemical and physical effects. Examples of the former include histone post-translational modifications, DNA methylation, and chromatin interactions with chromatin-binding proteins \cite{Allis2016, Maison2004, Madhani2018, Mango2018, Karpen2017, Larson2017, Larson2018}, while the latter include mechanical forces, geometric constraints, and hydrodynamic flows \cite{Spector2010, Lammerding2011, Zidovska2020b, Eshghi2021}. These interactions collectively govern the formation and maintenance of chromatin functional compartments, euchromatin and heterochromatin, while enabling their dynamical remodeling \cite{Karpen2018, Eshghi2021, Dekker2020, Zidovska2020b}. Understanding how these factors contribute to chromatin organization is essential for elucidating the mechanisms that regulate the genome's proper function \cite{Zidovska2020b, Misteli2020}.
	
	Adding to this complexity, chromatin is highly dynamic, being continually remodeled by molecular motors of transcriptional machinery, and other active processes \cite{Zidovska2013, Zidovska2024, Zidovska2020}. These active forces generate non-equilibrium stresses, introducing persistent fluctuations that alter the genome's  spatial organization \cite{Zidovska2013}. Despite significant advances in chromosome conformation capture and imaging techniques, which revealed new insights into chromatin organization and dynamics \cite{Volk2021, Gibcus2013, Liu2018}, the physical principles underlying the genome's spatial organization remain far from understood \cite{Misteli2020, Spector2010}. 
	
	Theoretical models offer a powerful framework to explore potential governing principles, providing insights into how chromatin dynamics, active stresses and mechanical forces interact to shape the genome's organization. To date, several different classes of chromatin models have been developed. Among them, continuum models provide a broad perspective by describing chromatin as a fluid subjected to active stresses and hydrodynamic interactions \cite{Bruinsma2014, Eshghi2021,Eshghi2023, Grosberg2023}. These models identified hydrodynamically interacting dipolar forces as a mechanism that can lead to large-scale coherent motions of the chromatin. Phase separation models describe chromatin compartmentalization through liquid-liquid or microphase separation, where segment-specific interactions and nuclear confinement drive euchromatin and heterochromatin segregation \cite{Mirny2019, Safran2021, Safran2023, Potoyan2024}. Polymer models extend this by incorporating cross-linking and epigenetic modifications, often assuming quasi-equilibrium conditions \cite{Menon2014, Marenduzzo2016, Kremer2017, Bloom2017, Spakowitz2018, Thirumalai2018, Onuchic2018, Spakowitz2020, Jost2021, Marko2024}. Active polymer models go beyond this framework by introducing non-equilibrium forces exerted on the chromatin fiber and ambient fluid by molecular motors such as RNA polymerase II, recapitulating chromatin coherent motions driven by ATP-dependent activity \cite{Saintillan2018, Saintillan2022, Shivashankar2022, Kardar2023, Weady2024}. 

    In this work, we develop a continuum framework that describes the large-scale organization of euchromatin and heterochromatin inside the cell nucleus.  {While earlier hydrodynamic models treated chromatin as a single active fluid \cite{Bruinsma2014, Eshghi2023, Grosberg2023}, our approach accounts for both euchromatin and heterochromatin as distinct compressible viscous fluids, which interact with the surrounding incompressible nucleoplasm. It also differs from polymer-based models \cite{Saintillan2018, Saintillan2022}, adopting here a coarser continuum description of chromatin organization.} Our model reveals that chromatin phase separation into euchromatin and heterochromatin is driven mechanically, facilitated by contractile stresses from the heterochromatin cross-linking proteins. This leads to the formation and growth of heterochromatin droplets,  {merging} into larger ones over time. Moreover, active processes in euchromatin regions (e.g., transcription) introduce persistent non-equilibrium fluctuations, which  {drive} the genome's spatial organization in the nucleus~\cite{Zidovska2013,Zidovska2024}. These fluctuations generate long-range, highly anisotropic structures within euchromatin, which in turn can deform the heterochromatin droplets. Unlike open systems, where phase separation  {produces} a single large droplet, nuclear confinement causes the heterochromatin to wet the nuclear boundary, forming a dynamic, shell-like layer. Notably,  {such peripheral heterochromatin organization} is ubiquitously observed in living cells~\cite{Bizhanova2021}.  {In our model, it arises directly from the boundary conditions imposed on chromatin fluids.} We find that the heterochromatin at the nuclear boundary is continuously deformed by active stresses  {acting through} euchromatin. Overall, our findings show how the interplay of active processes, mechanical forces and hydrodynamic flows shapes the genome's spatial organization in the cell nucleus.
	
	\vspace{-6pt}\section{Model and Equations of Motion}	
	
	We consider a spherical nucleus of radius $R$, with an undeformable nuclear envelope. The chromatin within the nucleus is treated as a continuous medium on spatial scales larger than the mesh size of its polymeric structure and on temporal scales longer than the stress relaxation time $\tau$ of the chromatin network \cite{Eshghi2021}. On deformation timescales much longer than $\tau$, the chromatin behaves as a viscous fluid. In contrast, on timescales comparable to or shorter than $\tau$, its rheological behavior becomes more complex, exhibiting nonlinear viscoelastic properties \cite{Eshghi2021}.  {In addition, ATP-dependent processes, such as transcription and chromatin remodeling, drive nonequilibrium positional fluctuations of chromatin; live-cell imaging reports micron-scale correlated chromatin motions that are eliminated upon ATP depletion \cite{Zidovska2013}. These observations are consistent with an effective fluidization of chromatin at mesoscopic scales \cite{Eshghi2021}. Within this regime, we model chromatin as an active viscous continuum hydrodynamically coupled to the nucleoplasmic solvent.}
	
\begin{figure}[t!]\includegraphics[width=\columnwidth]{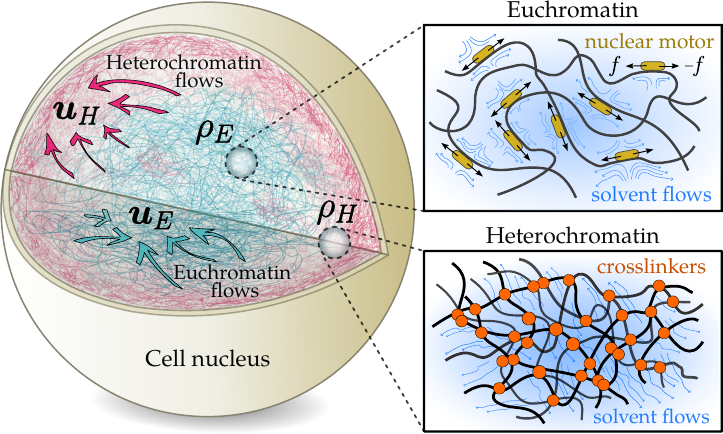}
		\caption{\label{fig:figures/figure_1}   Illustration of the cell nucleus, with chromatin compartmentalized into euchromatin (cyan) and heterochromatin (pink). We define the local density of euchromatin and heterochromatin by $\rho_E$ and $\rho_H$, respectively, and their corresponding velocity fields $\boldsymbol{u}_E$ and $\boldsymbol{u}_H$. At a microscopic level, euchromatin is loosely packed and transcriptionally active chromatin. Here, bound molecular motors to the euchromatin exert extensile stresses onto the neighboring aqueous environment, resulting in solvent flows (see inset). In contrast, heterochromatin consists mostly of transcriptionally silent genes, being densely packed by cross-linking molecules, which exert contractile stresses on the surrounding fluid, also resulting in solvent flows as shown in the inset.  
		}
	\end{figure}

	Beyond its polymeric nature, chromatin has a highly heterogeneous composition. In general, it has been categorized into two functional types: euchromatin and heterochromatin, which differ in their structure and biological functions \cite{Bonev2016}. Euchromatin is loosely packed and transcriptionally active, while heterochromatin is densely packed and transcriptionally silent, due to chromatin cross-linkers like heterochromatin protein 1 (HP1) and epigenetic modifications \cite{Karpen2018}. Since euchromatin and heterochromatin are defined by distinct components and microstructures, we model them as separate continuum media. These two media interact through local stresses, as well as frictional forces with the surrounding fluid (nucleoplasm), which are both mediated by the local chromatin composition. Here, the composition of each type is characterized by a local density of its respective components, distributed throughout the whole nuclear volume. We denote by $\rho_E$ and $\rho_H$ the number density of euchromatin and heterochromatin, respectively, as depicted in Fig.~\ref{fig:figures/figure_1}. 
	
	By  conservation of total euchromatin material in the nucleus, we have  
	\begin{equation}
		\label{eqn:kinem-E}
		\frac{\partial\rho_E}{\partial t}+\boldsymbol{\nabla}\cdot(\rho_E\,\boldsymbol{u}_E) = d_c\,\Delta\rho_E,
	\end{equation} where $\boldsymbol{u}_E(\boldsymbol{r},t)$ is the local euchromatin velocity. Similarly, the conservation of heterochromatin gives 
	\begin{equation}
		\label{eqn:kinem-H}
		\frac{\partial\rho_H}{\partial t}+\boldsymbol{\nabla}\cdot(\rho_H\,\boldsymbol{u}_H) = d_c\,\Delta\rho_H,
	\end{equation} where $\boldsymbol{u}_H(\boldsymbol{r},t)$ is the local velocity of heterochromatin. Here, the density of chromatin components evolves not only through diffusion but also through advection by their respective velocity fields. For simplicity, we assume that the diffusivity of both euchromatic and heterochromatic elements is governed by the same diffusion constant, $d_c$,  {a coarse-grained parameter describing random local rearrangements of chromatin density due to polymer conformational fluctuations and other small-scale mixing processes}. Given initial data for the densities $\rho_E$ and $\rho_H$, their evolution can be readily obtained once the velocities $\boldsymbol{u}_E$ and $\boldsymbol{u}_H$ are known. The latter are determined by the momentum equations associated with euchromatin and heterochromatin, respectively, which themselves depend explicitly on the chromatin densities. 
	
	The momentum equations are given by the balance of internal forces and friction with the surrounding solvent (nucleoplasm). Here, we neglect the inertial terms, assuming that their dynamics is dominated by the frictional forces with the solvent. The momentum conservation equation for the euchromatin is given by
	\begin{equation}
		\label{eqn:stress-E-balance}
		\boldsymbol{\nabla}\cdot\boldsymbol{\sigma}_E = \Gamma_0\hspace{1pt}\rho_E\left(\boldsymbol{u}_E-\boldsymbol{u}_S\right)\!,
	\end{equation} where $\Gamma_0$ is a  {drag coefficient that controls the strength of the friction force with the solvent}, $\boldsymbol{u}_S(\boldsymbol{r},t)$ is the solvent velocity, and $\boldsymbol{\sigma}_E$ is the euchromatin stress tensor, which is given by the following constitutive relation \cite{Kirkwood1958, Snell1967, Onuki1992,Doi2013}:
	\begin{equation}
		\label{eqn:euch-stress}
		\boldsymbol{\sigma}_E = {-p_E\hspace{1pt}\boldsymbol{I} + 2\eta_c\rho_E\boldsymbol{E}_E}
	\end{equation} where {$\eta_c\rho_E$} is the density-dependent viscosity of euchromatin, with $\eta_c$ a viscosity coefficient, $\boldsymbol{E}_E$ is the rate-of-strain tensor of the euchromatin material, that is,  
	\begin{equation}
		\boldsymbol{E}_E =\frac{1}{2} [\boldsymbol{\nabla}\boldsymbol{u}_E+(\boldsymbol{\nabla}\boldsymbol{u}_E)^{\!\mathsf{T}}],
	\end{equation} and the euchromatic pressure is given by
	\begin{equation}
		\label{eqn:pressure-E}
		{p_E = \frac{1}{2}B\rho_E(\rho_E+\rho_H)},
	\end{equation} where $B$ captures the strength of steric interactions between euchromatin and total chromatin. {Note that we do not subtract the trace of the strain tensor, and thus include both shear and volumetric contributions. This corresponds to a two-constant approximation, using a single coefficient for both shear and dilatational viscosity.}  
	\\
	\indent
	Similarly, we write the force balance equation for the heterochromatin to be
	\begin{equation}
		\label{eqn:stress-H-balance}
		\boldsymbol{\nabla}\cdot\boldsymbol{\sigma}_H =\Gamma_0\hspace{1pt}\rho_H(\boldsymbol{u}_H-\boldsymbol{u}_S),
	\end{equation} where the heterochromatin stress $\boldsymbol{\sigma}_H$ is given by
	\begin{equation}
		\label{eqn:hetero-stress}
		\boldsymbol{\sigma}_H = {-p_H\boldsymbol{I}+2\eta_c\rho_H\boldsymbol{E}_H},	
	\end{equation} with {$\eta_c\rho_H$} as the heterochromatin density-dependent viscosity, and the rate-of-strain tensor 
	\begin{equation}
		\boldsymbol{E}_H =\frac{1}{2} [\boldsymbol{\nabla}\boldsymbol{u}_H+(\boldsymbol{\nabla}\boldsymbol{u}_H)^{\!\mathsf{T}}].
	\end{equation} The heterochromatic pressure $p_H$ is given by~\cite{Foster2015, Toner2005}:
	\begin{equation}
		\label{eqn:pressure-H}
		{p_H = \frac{1}{2}B\rho_H(\rho_E+\rho_H) - B {\rho}_C\rho_H},
	\end{equation}  where $B$ is the same as in Eq.~(\ref{eqn:pressure-E}),  {and contractility $\rho_C$ parameterizes an \emph{isotropic contractile} stress that favors heterochromatin condensation and opposes steric repulsion. While HP1 oligomerization mediates \emph{passive} condensation \cite{Larson2017}, the establishment of heterochromatin also depends on ATP-consuming processes. These include events during differentiation, when repressive histone marks such as H3K9 (together with other epigenetic modifications) are deposited, as well as transitions of facultative heterochromatin, when genetic loci are silenced/reactivated. Specific mechanisms include reader--writer feedback linking H3K9 methyltransferases such as SUV39H1/2 to HP1 binding at H3K9me2/3 marks~\cite{Lachner2001,Padeken2022}, and more broadly ATP-dependent chromatin remodelers that restructure nucleosomes and regulate accessibility~\cite{Clapier2017}. HP1 has low affinity for unmarked chromatin, binding strongly after H3K9 methylation~\cite{Lachner2001,Padeken2022}. Thus, $\rho_C$ is an \emph{effective compaction parameter} that incorporates both the history of these energy-dependent contributions and the passive condensation by HP1. Physically, $\rho_C$ sets the density threshold beyond which contractile stresses dominate over steric repulsion~\cite{Foster2015}.}
	\begin{table*}
		\begin{ruledtabular}
			\begin{tabular}{c|cc|c}
				&Continuity equation: & 
				& Momentum balance equation:\\[3pt]\hline
				& & &\\[-10pt]
				SL: &$\displaystyle\boldsymbol{\nabla}\cdot\boldsymbol{u}_S = 0$ & 
				& $\displaystyle-\boldsymbol{\nabla} p + \Delta\boldsymbol{u}_S\hspace{1pt}-\boldsymbol{\nabla} \cdot (\alpha\rho_E\boldsymbol{Q})  =  \Gamma\!\left[\rho_H(\boldsymbol{u}_S-\boldsymbol{u}_H)+\rho_E(\boldsymbol{u}_S-\boldsymbol{u}_E)\right]$\\[14pt]
				EC: 
				&$\displaystyle\frac{\partial\rho_E}{\partial t}+\boldsymbol{\nabla}\cdot(\rho_E\,\boldsymbol{u}_E) = \Delta\rho_E$ & 
				&$\displaystyle-\frac{1}{2}\boldsymbol{\nabla}[\rho_E(\rho_E+\rho_H)]+{\boldsymbol{\nabla}\cdot\big[\eta\,\rho_E\big(\boldsymbol{\nabla}\boldsymbol{u}_E+(\boldsymbol{\nabla}\boldsymbol{u}_E)^\mathsf{T}\big)\big]} = \Gamma \rho_E(\boldsymbol{u}_E-\boldsymbol{u}_S)$\\[16pt]
				HC:\; 
				&$\displaystyle\frac{\partial\rho_H}{\partial t}+\boldsymbol{\nabla}\cdot(\rho_H\,\boldsymbol{u}_H) = \Delta\rho_H$ & 
				&$\displaystyle-\frac{1}{2}\boldsymbol{\nabla}[\rho_H(\rho_E+\rho_H)-2\bar{\rho}_C\rho_H]+{\boldsymbol{\nabla}\cdot\big[\eta\,\rho_H\big(\boldsymbol{\nabla}\boldsymbol{u}_H+(\boldsymbol{\nabla}\boldsymbol{u}_H)^\mathsf{T}\big)\big]}  = \Gamma \rho_H(\boldsymbol{u}_H-\boldsymbol{u}_S)$\\[12pt]
				\hline\hline
				& Order parameter conditions: & 
				& Dynamics of nematic order parameter:\\[2pt]\hline
				& & &\\[-10pt]
				&$\displaystyle\boldsymbol{Q}=\boldsymbol{Q}^{\mathsf{T}}$ and $\;\displaystyle\mathrm{tr}(\boldsymbol{Q})=1$& 
				&$\displaystyle\frac{\partial\boldsymbol{Q}}{\partial t}+\boldsymbol{u}_S\cdot\boldsymbol{\nabla}\boldsymbol{Q}-\boldsymbol{Q}\cdot(\boldsymbol{\nabla}\boldsymbol{u}_S)-(\boldsymbol{\nabla}\boldsymbol{u}_S)^{\mathsf{T}}\cdot\boldsymbol{Q} =  \Delta\boldsymbol{Q}-2\hspace{1pt}\boldsymbol{Q}\,\boldsymbol{Q}:\boldsymbol{\nabla}\boldsymbol{u}_S$\\[5pt]
			\end{tabular}
		\end{ruledtabular}
		\caption{\label{tab:scaled-eqs} Dimensionless equations for the number conservation of solvent (SL), euchromatin (EC) and heterochromatin (HC) components, and their corresponding dimensionless momentum balance equations. Also, we include the dimensionless dynamical equation of the nematic order parameter $\boldsymbol{Q}$, conditioned to be a symmetric tensor with trace one.}
	\end{table*}
	\\
	\indent
	We assume that euchromatin and heterochromatin are dilute phases within the ambient solvent (nucleoplasm). Thus, we treat the solvent as an incompressible viscous fluid satisfying the condition:
	\begin{equation}
		\label{eqn:imp-cond}
		\boldsymbol{\nabla}\cdot\boldsymbol{u}_S = 0.
	\end{equation}  {Although chromatin occupies an appreciable fraction of nuclear volume (about 10--30\%), 
    it forms a deformable, solvent-permeated network rather than a dense medium. From a coarse-grained hydrodynamic perspective, the nucleoplasm remains the dominant carrier of volume, and thus treating chromatin components as dilute dispersed phases is a reasonable first approximation. The effect of crowding is captured by the chromatin viscosity coefficient $\eta_c$ and drag coefficient $\Gamma_0$, which define a screening length $\ell_{c}=(\eta_c/\Gamma_0)^{1/2}$. This length sets how far disturbances are transmitted through the chromatin before being damped by drag against the solvent. At distances shorter than $\ell_c$, chromatin viscosity dominates and disturbances spread in a Stokes-like fashion. At larger distances, friction dominates, relative velocities are suppressed, and the chromatin phases move together with the solvent ($\mathbf u_E,\mathbf u_H \approx \mathbf u_S$). In the limiting cases, $\Gamma_0 \to 0$ ($\ell_c\to\infty$) corresponds to unscreened propagation across the entire nucleus, while $\Gamma_0 \to \infty$ ($\ell_c\to 0$) corresponds to complete locking of chromatin to the solvent.  }
	\\
	\indent
    The force balance equation of the solvent is given by 
	\begin{equation}
		\label{eqn:stokes-solvent}
		\!\!\boldsymbol{\nabla}\cdot\boldsymbol{\sigma}_S = \Gamma_0\hspace{1pt}\rho_H(\boldsymbol{u}_S-\boldsymbol{u}_H)+\Gamma_0\hspace{1pt}\rho_E(\boldsymbol{u}_S-\boldsymbol{u}_E),
	\end{equation} with the solvent stress tensor 
	\begin{equation}
		\label{eqn:solvent-stress}
		\boldsymbol{\sigma}_S = -p_S\boldsymbol{I}+2\eta_s\boldsymbol{E}_S + \boldsymbol{\sigma}_A,
	\end{equation} where $p_S$ is the solvent pressure, $\eta_s$ is the viscosity of the solvent, and $\boldsymbol{E}_S$ is the rate-of-strain tensor,
	\begin{equation}
		\boldsymbol{E}_S =\frac{1}{2} [\boldsymbol{\nabla}\boldsymbol{u}_S+(\boldsymbol{\nabla}\boldsymbol{u}_S)^{\!\mathsf{T}}].
	\end{equation} Here the solvent pressure $p_S$ acts as a Lagrange multiplier that enforces the incompressibility condition in Eq.~(\ref{eqn:imp-cond}). Similarly, in deriving Eq.~(\ref{eqn:stokes-solvent}), we neglect the inertial terms, assuming that friction is the dominant relaxation mechanism. In addition to the Newtonian stress given by the first two terms on the right-hand-side of Eq.~(\ref{eqn:solvent-stress}), the solvent is also driven by an active stress, $\boldsymbol{\sigma}_A$,  {arising from ATP-dependent nuclear enzymes bound to euchromatin:}
	\begin{equation}
		\boldsymbol{\sigma}_A = -\alpha_0\rho_E\hspace{0.5pt}\boldsymbol{Q},
	\end{equation}  {where $\alpha_0$ is the coupling strength associated with enzyme activity, {\it negative} for contractile stresses and {\it positive} for extensile stresses. The active stress} depends on the local density of euchromatin and on a nematic order parameter $\boldsymbol{Q}(\boldsymbol{r},t)$, which is a second-rank tensor that characterizes the alignment of  {dipolar forces exerted onto the solvent by nuclear enzymes (e.g., transcriptional machinery, and chromatin remodelers) associated with euchromatin}~\cite{Saintillan2018}. Specifically, $\boldsymbol{Q}$ is a symmetric tensor, $\boldsymbol{Q} = \boldsymbol{Q}^{\mathsf{T}}$, with trace $\mathrm{tr}(\boldsymbol{Q})=1$, and it satisfies~\cite{Saintillan2008, Saintillan2013, Gao2017[Rev]}:  
	\begin{equation}
		\label{eqn:Q-dyanmics}
		\overset{\triangledown}{\boldsymbol{Q}} = d_c\hspace{1pt}\Delta\boldsymbol{Q}-2\hspace{1pt}\boldsymbol{Q}\,\boldsymbol{Q}:\boldsymbol{E}_S
	\end{equation} where $\overset{\triangledown}{}$ denotes the upper convective derivative,
	\begin{equation}
		\overset{\triangledown}{\boldsymbol{Q}} = \frac{\partial\boldsymbol{Q}}{\partial t}+\boldsymbol{u}_S\cdot\boldsymbol{\nabla}\boldsymbol{Q}-\boldsymbol{Q}\cdot(\boldsymbol{\nabla}\boldsymbol{u}_S)-(\boldsymbol{\nabla}\boldsymbol{u}_S)^{\mathsf{T}}\cdot\boldsymbol{Q}.
	\end{equation}   {At the microscopic level, the active forces exerted onto the solvent originate from enzymes bound along segments of euchromatin. Each such segment, with its bound nuclear motors, acts as a force dipole~\cite{Bruinsma2014,Saintillan2018}.} Eq.~(\ref{eqn:Q-dyanmics}) is derived from a kinetic theory which describes the spatial and orientational distribution of  {these active dipolar units}~\cite{Saintillan2008, Saintillan2013, Gao2017[Rev]}.  {These units are modeled as geometrically anisotropic, rod-like structures that are advected, rotated, and aligned by the solvent flow, and undergo translational diffusion with a diffusivity taken to be the same as the chromatin diffusivity $d_c$.} By averaging over the orientational degrees of freedom,  {the kinetic equation for their} distribution can be  {expressed} in terms of orientational moments.  {The resulting equation for $\boldsymbol{Q}$ then involves the fourth-order moment, which we approximate by $\boldsymbol{Q}\boldsymbol{Q}$ (quadratic closure)~\cite{DoiBook1986} to obtain Eq.~(\ref{eqn:Q-dyanmics}). 
    Other closures may differ in detail but generally lead to the same qualitative dynamics~\cite{Maddu2024}. 
    Although the actual microscopic force distribution along chromatin segments is likely more complex, the present description provides a minimal anisotropic model that captures fiber-aligned activity and couples local force generation to solvent flow. This leads to a long-wavelength instability~\cite{Saintillan2013} and, as shown below, is consistent with nucleus-scale displacement correlations observed in experiments~\cite{Zidovska2013}.} 
	\\
	\indent
	Note that the overall system is force free. Indeed, combining Eqs.~(\ref{eqn:stress-E-balance}), (\ref{eqn:stress-H-balance}), and (\ref{eqn:stokes-solvent}) leads to
	\begin{equation}   \boldsymbol{\nabla}\cdot\left[\,\boldsymbol{\sigma}_S+\boldsymbol{\sigma}_E+\boldsymbol{\sigma}_H\right] = 0.
	\end{equation} {Note also that by eliminating heterochromatin contractility (${\rho}_C = 0 $) and euchromatin activity ($ \alpha = 0 $), and then assuming a common velocity for the two chromatin phases, $ \boldsymbol{u}_E = \boldsymbol{u}_H = \boldsymbol{u} $, we obtain a consistent single-phase description of chromatin with velocity $ \boldsymbol{u} $ and density $ \rho_E + \rho_H $ by adding together the respective momentum and number conservation equations. In this limit, the nematic order parameter equation decouples completely from the chromatin momentum equations.}
	\\
	\indent
     {The governing equations are nondimensionalized, as summarized in Table I, using the nucleus radius $R$ as the length scale, the diffusion time $T = R^2/d_c$ as the time scale, and a characteristic chromatin density scale $\rho_0 = \sqrt{(\eta_s d_c)/(R^2 B)}$ as derived in Appendix~A. This leads to only four independent dimensionless parameters: the activity parameter $\alpha = \alpha_0 R/\sqrt{\eta_s d_c B}$, the reduced viscosity $\eta=\eta_c\sqrt{d_c/(\eta_s R^2 B)}$, the friction coefficient $\Gamma = \Gamma_0 R\sqrt{d_c/(\eta_s B)}$, and the contractility parameter $\bar{\rho}_C = \rho_C R \sqrt{B/(\eta_s d_c)}$. Parameter estimates and their physiological ranges are discussed in Appendix~B, with dimensional values summarized in Table II.}
	
	\vspace{-6pt}\section{Heterochromatin coarsening}

\begin{figure*}[t!]\includegraphics[width=\textwidth]{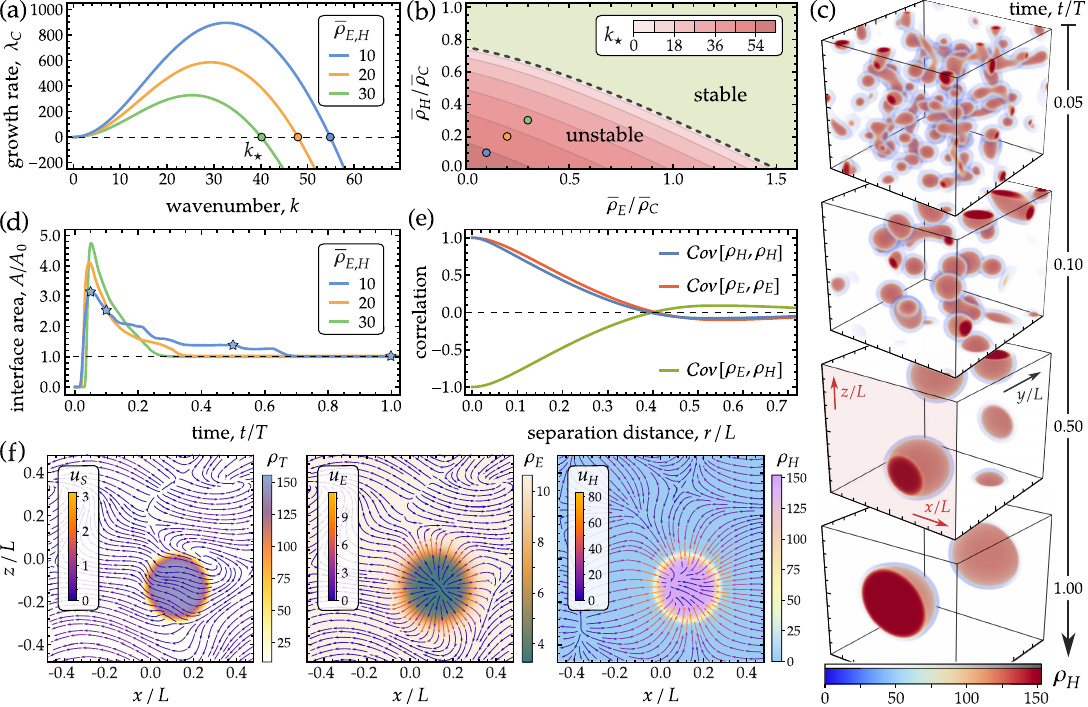}
		\caption{\label{fig:figures/figure_2} (a)~Growth rates $\lambda_C$ of density perturbations for each Fourier mode $k$, perturbed around  the homogeneous densities $\bar{\rho}_E$ and $\bar{\rho}_H$; here, we choose $\bar{\rho}_E=\bar{\rho}_H$. (b)~Linearly unstable (red) and stable (green)  regions, where the stability boundary is shown by the dashed curve. Contour lines in the unstable region indicate the largest unstable mode $k_\star$, while the three points correspond to those in (a) using the same color convention.  {(c)~Density snapshots of the heterochromatin in a three-dimensional periodic domain of equal lengths $L$, by initially applying small density perturbations about the base state {$\bar{\rho}_{E,H}=10$}. Only high-density regions are shown (see color bar); low-density regions are rendered transparent, with opacity indicated by the grayscale bar. The instability leads to the formation of small heterochromatic droplets, which subsequently coalesce into a single spherical droplet at long times. The red-highlighted slice indicates the two-dimensional plane used in panel~(f), and the overlaid arrows indicate the~$x$ and~$y$ directions. All coordinates are nondimensionalized by the box size~$L$, which corresponds to the nuclear diameter.} (d)~Total interfacial area $A$ of droplets, rescaled by the effective area $A_0 =\sqrt[3]{36\pi }\,V_0^{2/3}$, where $V_0$ is the volume of all droplets at time $t$. The {blue stars} correspond to the snapshots in (c). At long times, $A/A_0$ relaxes to one; that is, a single spherical droplet. (e)~Two-point correlations of density fluctuations from their mean, normalized by their standard deviations, at a radial separation distance $r$, where the average is taken over all possible points separated by $r$ and all snapshots with $t/T\geq{0.20}$. The size of shaded regions corresponds to the standard deviation of the sample.  {(f) Densities $\rho_E$ and $\rho_H$, and total density $\rho_T = \rho_E+\rho_H$,  shown on the two-dimensional slice highlighted in red in panel (c). Streamlines represent the projected velocity fields $\boldsymbol{u}_S$, $\boldsymbol{u}_E$, and $\boldsymbol{u}_H$ on this slice.} The streamlines are colored by the magnitude of the velocity, as shown by the insets, whereas the background color indicates the density. Herein, we use $\eta=0.01$, $\bar{\rho}_C = 100$, and $\Gamma=25$.}
	\end{figure*}

         {To understand the dynamics of the system (summarized in dimensionless form in Table I)}, we first consider the activity strength $\alpha=0$, and study the dynamics only in the presence of contractile stresses due to heterochromatin cross-linking ($\bar{\rho}_C\neq0$). We restrict our system to a three-dimensional periodic domain of {equal side lengths $L=2R$,  {corresponding to the nuclear diameter,} and volume $V = L^3$}. This allows us to study the linear stability of the system about its stationary and homogeneous state, which is given by $\boldsymbol{u}_E=\boldsymbol{u}_H=\boldsymbol{u}_S=0$, and $\rho_E = \bar{\rho}_{E}$ and $\rho_H = \bar{\rho}_{H}$, where $\bar{\rho}_{E}$ and $\bar{\rho}_{H}$ are constant densities, which correspond to the overall number of euchromatin and heterochromatin components in the system (i.e.~$N_E = \bar{\rho}_{E} V$ and $N_H = \bar{\rho}_{H} V$, respectively).
	\\
	\indent
	We perturb the system around the homogeneous state by introducing small density fluctuations in euchromatin and heterochromatin; that is,
	$\rho_E(\boldsymbol{r},t) = \bar{\rho}_E+\delta\rho_E(\boldsymbol{r},t)$ and $\rho_H(\boldsymbol{r},t) = \bar{\rho}_H+\delta\rho_H(\boldsymbol{r},t)$, with $\delta{\rho}_E$ and $\delta{\rho}_H$ as the density perturbations. This also induces perturbations in the velocity fields of euchromatin, heterochromatin, and  solvent, which we denote by $\delta\boldsymbol{u}_E$, $\delta\boldsymbol{u}_H$, and $\delta\boldsymbol{u}_S$, respectively. The linearized equations associated with these density and velocity perturbations can be derived as shown in  {Appendix~C}. In particular, the spatiotemporal evolution of density perturbations can be reduced to two coupled linear partial differential equations in terms of only $\delta\rho_E$ and $\delta\rho_H$ (see  {Appendix~C.1}). This is achieved by explicitly eliminating their dependence on the velocity perturbations using the linearized momentum equations of euchromatin and heterochromatin, together with the incompressibility condition of the solvent in Eq.~(\ref{eqn:imp-cond}). Since we consider a periodic domain, these linear equations can be decomposed into Fourier modes; namely, $\delta\rho_{E,H}(\boldsymbol{r},t) = \sum_{\boldsymbol{k}}\,e^{i\boldsymbol{k}\cdot\boldsymbol{r}}\,\delta\hat{\rho}_{E,H}^{\,\boldsymbol{k}}(t)$, where $\delta\hat{\rho}_{E}^{\,\boldsymbol{k}}$ and $\delta\hat{\rho}_{H}^{\,\boldsymbol{k}}$ are the density amplitudes associated with the Fourier wave-vector~$\boldsymbol{k}$. By orthogonality, each amplitude at a specific frequency is independent of the amplitudes at other frequencies, and satisfies the following equation:
	\begin{equation}
		\frac{\partial}{\partial t}\!\begin{bmatrix}
			\delta\hat{\rho}_{E}^{\,\boldsymbol{k}} \\[3pt]
			\delta\hat{\rho}_{H}^{\,\boldsymbol{k}} 
		\end{bmatrix} =	 \,\boldsymbol{\mathcal{S}}_{\hspace{-0.5pt}{k}}\!\begin{bmatrix}\delta\hat{\rho}_{E}^{\,\boldsymbol{k}} \\[3pt]
			\delta\hat{\rho}_{H}^{\,\boldsymbol{k}} 
		\end{bmatrix}\!,
	\end{equation}
    where
	\begin{equation}
		\label{eqn:stability-matrix}
		\boldsymbol{\mathcal{S}}_{\hspace{-0.5pt}{k}} = -k^2\!
		\begin{bmatrix}
			\displaystyle\, 1+\frac{\bar{\rho}_E + \frac{1}{2}\bar{\rho}_H}{\Gamma+{2 \eta}  k^2}
			& \displaystyle \frac{\frac{1}{2}\bar{\rho}_E}{\Gamma +{2 \eta}  k^2} \\[12pt]
			\displaystyle \frac{\frac{1}{2}\bar{\rho}_H}{\Gamma +{2 \eta}  k^2}
			& \displaystyle 1+\frac{\bar{\rho}_H - \bar{\rho}_C + \frac{1}{2}\bar{\rho}_E}{\Gamma+{2 \eta} k^2}\,
		\end{bmatrix}\!
	\end{equation} and $k$ is the magnitude of $\boldsymbol{k}$. The eigenvalues of $\boldsymbol{\mathcal{S}}_{\hspace{-0.5pt}{k}}$ correspond to the growth rates of the Fourier amplitudes of density perturbations. We find that one of the eigenvalues is always negative, while the other eigenvalue that we denote by $\lambda_C$ can become positive over a finite band of Fourier modes, as shown in Fig.~\ref{fig:figures/figure_2}(a). This means that the Fourier modes with $k<k_\star$ are linearly unstable. The largest unstable mode $k_\star$ can be used to determine the region of instability ($k_\star>0$) and the stability boundary ($k_\star=0$) by varying the constant densities $\bar{\rho}_E$ and $\bar{\rho}_H$ at fixed $\Gamma$ and $\bar{\rho}_C$, see Fig.~\ref{fig:figures/figure_2}(b). The instability condition can be found in exact form ( {Appendix C.1}):
	\begin{equation}
		\label{eqn:condition-instability}
		\bar{\rho}_C> \bar{\rho}^{\,\star}_C =\frac{(\Gamma + \bar{\rho}_E+\bar{\rho}_H)(2\Gamma + \bar{\rho}_E+\bar{\rho}_H)}{2(\Gamma+\bar{\rho}_E)+\bar{\rho}_H}.
	\end{equation}
	Thus, the system is unstable if the contractility $\bar{\rho}_C$ is greater than a threshold $\bar{\rho}^{\,\star}_C$, corresponding to a regime in which the contractile forces due to heterochromatin cross-linking dominate over the frictional forces. 

    In the linearly unstable region, the system is driven away from the homogeneous state. To study the behavior far from the linear regime, we solve numerically the governing equations by initially applying a density perturbation in euchromatin and heterochromatin about their homogeneous densities. Here, we choose initial density perturbations such that $\int\delta\rho_{E,H}(\boldsymbol{r},0)\,\mathrm{d}^3\boldsymbol{r}=0$, leaving the total number of euchromatin and heterochromatin components unchanged, with $N_{E,H} = \bar{\rho}_{E,H}V$. We employ a fast spectral method (Dedalus) with Fourier domain discretization in three dimensions and an implicit-explicit fourth-order Runge-Kutta scheme for time stepping~\cite{Dedalus}. Selecting homogeneous densities $\bar{\rho}_E$ and $\bar{\rho}_H$, along with parameters $\eta$, $\Gamma$, and $\bar{\rho}_C$, such that the system is linearly unstable, Fig.~\ref{fig:figures/figure_2}(c) presents snapshots of heterochromatin density from a numerical simulation, corresponding to  {Video~1} in the  {Supplementary Material~(SM)~\footnote{ {See Supplemental Material for additional figures and videos.}}}, {with density $\bar{\rho}_{E,H} = 10$}. This reveals how the instability leads to the coarsening of heterochromatin, beginning with the formation of small droplets that grow and merge over time, ultimately resulting in a single spherical droplet at long times. At higher concentrations, the finite size of the periodic domain significantly influences the long-term morphology of heterochromatin. For instance, at {$\bar{\rho}_{E,H} = 20$ (see  {Video~2})}, heterochromatin forms an elongated cylindrical structure that spans the domain in one direction, maintaining a radius smaller than $L$. At even higher $\bar{\rho}_H$, heterochromatin adopts a planar configuration spanning two dimensions of the domain, with a finite thickness less than $L$, {see  {Video~3}, where $\bar{\rho}_{E,H} = 40$}.
    \\
	\indent
	Since heterochromatin rapidly reaches its maximum equilibrium density, $\rho_{H}^{\mathrm{max}}$, and forms sharp interfaces between euchromatin-rich and heterochromatin-rich regions, see Fig.~\ref{fig:figures/figure_2}(c), we identify each droplet by its interfacial surface. Specifically, we define the interface as the region where the heterochromatin density $\rho_H$ equals to half of its maximum value, $\rho_{H} = \frac{1}{2}\rho_{H}^{\mathrm{max}}$. By triangulating this surface, the area and enclosed volume of each distinct droplet can be obtained. Fig.~\ref{fig:figures/figure_2}(d) shows the total interface area as a function of time, normalized by the effective spherical area $A_0 = \sqrt[3]{36\pi}\,V_0^{2/3}$, where $V_0$ is the total volume of heterochromatin droplets at time $t$. For cases where heterochromatin forms a single spherical droplet at long times, $A/A_0$ relaxes to one, while for cylindrical and planar morphologies, it stabilizes slightly above one due to finite-size effects. In all cases, the interface behavior reflects an effective surface tension emerging from the model. The reduced viscosity $\eta$ primarily controls the interface thickness, as the viscous terms in the momentum equations (see Table I) dominate at the interface, where density and velocity gradients are the steepest, thereby regulating the interface structure.
    \\
	\indent
	Fig.~\ref{fig:figures/figure_2}(f) shows the two-dimensional projection of the velocity fields of solvent, euchromatin, and heterochromatin, at a single snapshot in time (in Fig.~\ref{fig:figures/figure_2}(c), this projection slice is highlighted in red).  {Additional slices can be found in SM (see Fig.~S.1)}. The flows of heterochromatin are directed towards the high density regions of heterochromatin, whereas euchromatin flows out of these high density regions, being depleted within the heterochromatin droplets. This is due to the steric repulsion terms, which together with the viscous and contractile stresses result in a maximum steady-state value for the total density $\rho_T$ of chromatin, as shown in Fig.~\ref{fig:figures/figure_2}(f). 
	\begin{figure*}[t!]\includegraphics[width=\textwidth]{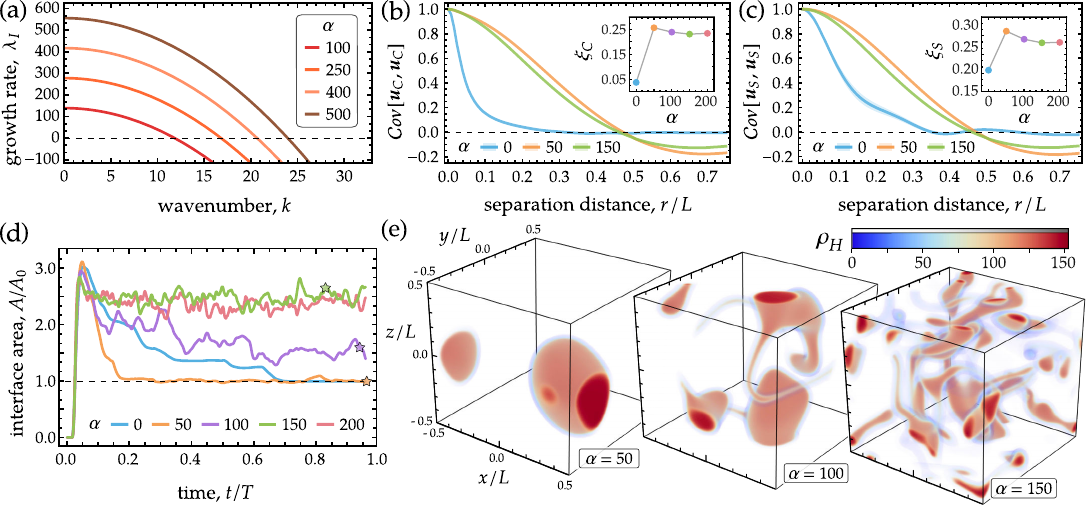}
		\caption{\label{fig:figures/figure_3} (a) Growth rates $\lambda_I$ associated with perturbations in the nematic order parameter $\boldsymbol{Q}$ about the isotropic state $\frac{1}{3}\boldsymbol{I}$ for different values of the activity strength $\alpha$. (b) Two-point correlation of the solvent velocity $\boldsymbol{u}_S$ at a separation $r$, as a function of $\alpha$, normalized by the variance, and averaged over all snapshots with $t/T\geq0.25$. Correlation length $\xi_S$ in units of the system size $L$ is shown in the inset plot. (c) Two-point correlation of the barycentric velocity of the chromatin $\boldsymbol{u}_C = (\rho_E \boldsymbol{u}_E+\rho_H \boldsymbol{u}_H)/\rho_T$ at a separation distance $r$, normalized by the variance, and averaged over all snapshots with $t/T\geq{0.20}$. Its associated correlation length $\xi_C$ (in units of $L$) is shown in the inset. In both (b) and (c), the size of shaded regions corresponds to the standard deviation of the sample. (d) Net interface area $A$ in terms of the spherical area $A_0$ as in Fig.~\ref{fig:figures/figure_2}(d) for different $\alpha$. For nonzero activity, the ratio $A/A_0$ fluctuates about some mean value {which increases as we increase $\alpha$, since at large activity $\alpha$} the shape of the droplets is continuously deformed and occasionally split into smaller droplets. (e) Heterochromatin density for different values of $\alpha$ {at snapshots indicated by the stars in (d)}.  {The color and opacity scheme and axis conventions are the same as in Fig.~2(c)}. Herein, {$\eta=0.01$, $\bar{\rho}_C = 100$, $\Gamma=25$, and $\bar{\rho}_{E,H}=10$}. 
		}
	\end{figure*}
    \\
	\indent
    The demixing between euchromatin and heterochromatin can be quantified by studying the two-point correlations of density fluctuations from the mean, measured at a fixed radial separation distance $r$; namely,
	\begin{widetext}
		\begin{equation}
			\label{eqn:cov_rhoE_rho_H}
			\textit{Cov}\left[\rho_E,\rho_H\right] = \frac{\iint\left(\rho_E(\boldsymbol{r}',t)-\bar{\rho}_E\right)\left(\rho_H(\boldsymbol{r}'',t)-\bar{\rho}_H\right)\delta(r-|\boldsymbol{r}'-\boldsymbol{r}''|)\,\mathrm{d}^3\boldsymbol{r}'\,\mathrm{d}^3\boldsymbol{r}''}{\frac{1}{V}\!\left|\int\left(\rho_E(\boldsymbol{r},t)-\bar{\rho}_E\right)\left(\rho_H(\boldsymbol{r},t)-\bar{\rho}_H\right)\mathrm{d}^3\boldsymbol{r}\right|\,\iint\delta(r-|\boldsymbol{r}'-\boldsymbol{r}''|)\,\mathrm{d}^3\boldsymbol{r}'\,\mathrm{d}^3\boldsymbol{r}''}.
		\end{equation}
	\end{widetext} In Fig.~\ref{fig:figures/figure_2}(e), the correlation in Eq.~(\ref{eqn:cov_rhoE_rho_H}) is shown by averaging over all simulation snapshots, ignoring the early dynamics (chosen here as $t/T\geq0.2$). This shows a negative correlation between euchromatin and heterochromatin densities at short-range separations, because of the chromatin demixing, which becomes positively correlated at a separation commensurate with the effective size of the  spherical droplet of heterochromatin at long times. Similarly, we compute the correlations corresponding to heterochromatin densities, $\textit{Cov}\left[\rho_H,\rho_H\right]$, and those associated with euchromatin densities, $\textit{Cov}\left[\rho_E,\rho_E\right]$, which we define in an analogous way to Eq.~(\ref{eqn:cov_rhoE_rho_H}). A time average of these correlations is shown in Fig.~\ref{fig:figures/figure_2}(e), which are both positively correlated on short-range separations. 
	
	\vspace{-6pt}\section{Euchromatin activity}
	
	We now examine the case of nonzero euchromatin activity strength $\alpha$, focusing on positive values that correspond to extensile active stresses, while keeping the same three-dimensional periodic domain of side length $L$. In addition to the instability driven by contractile stresses (controlled by $\bar{\rho}_C$), this activity  introduces a distinct linear instability governed by the activity parameter $\alpha$.
	\begin{figure*}[t!]\includegraphics[width=\textwidth]{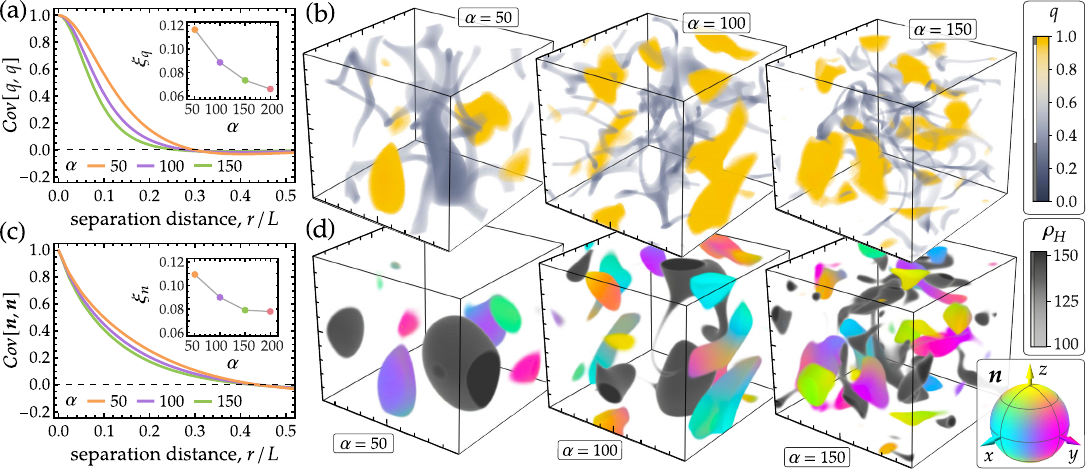}
		\caption{\label{fig:figures/figure_4} (a) Two-point correlation of fluctuations from the mean of the scalar order parameter {$q$}, at a radial separation $r$, normalized by the variance. Inset shows {the correlation length $\xi_q$ in the scalar order parameter $q$ at increasing values of $\alpha$}. (b) Three-dimensional plot of {$q$} for different $\alpha$, where only the high values (${q}\geq0.96$) and low values (${q}\leq0.36$) are opaque, while the rest is transparent; see color bar and the corresponding opacity levels. (c) Two-point correlation in fluctuations of the director field $\boldsymbol{n}$ around its mean, at a radial separation $r$, and normalized by the variance. Inset shows {the associated correlation length $\xi_n$ of the director field}. (d) The highly aligned regions in (b), shown in yellow, are instead colored by the orientation of the director field $\boldsymbol{n}$, as shown by the inset legend. By interpolating between three colors (cyan, magenta, and yellow), antipodal points on the unit sphere are associated with the same color (e.g.~pure yellow corresponds to both north and south pole). The dark gray  display the high density regions of heterochromatin, as in Fig.~\ref{fig:figures/figure_3}(e); see color bar of $\rho_H$ and opacity levels.  {In both (b) and (d), the same axis convention as in Fig.~3(e) is used.} Here, {we use $\eta=0.01$, $\bar{\rho}_C = 100$, $\Gamma=25$, and $\bar{\rho}_{E,H}=10$}.      
		}
	\end{figure*}
	\\
	\indent
	By perturbing the nematic order parameter about its isotropic state, that is, $\boldsymbol{Q}(\boldsymbol{r},t) = \frac{1}{3}\boldsymbol{I}+\delta\boldsymbol{Q}(\boldsymbol{r},t)$, its linearized dynamics is found to be ( {Appendix C.2}):
	\begin{equation}
		\label{eqn:lin-Q-dyn-iso}
		\frac{\partial}{\partial t}(\delta\boldsymbol{Q})=\frac{2}{3}\delta\boldsymbol{E}_S + \Delta(\delta\boldsymbol{Q}),
	\end{equation} where $\delta\boldsymbol{E}_S$ is the linearized rate of strain corresponding to a velocity perturbation $\delta\boldsymbol{u}_S$ in the solvent. Since $\boldsymbol{Q}$ is a symmetric tensor of trace one (see Table I), we must require $\delta\boldsymbol{Q}$ to be a symmetric and traceless tensor. Eq.~(\ref{eqn:lin-Q-dyn-iso}) can be rewritten in terms of the Fourier amplitude $\delta\boldsymbol{Q}_{\boldsymbol{k}}$ associated with the wave-vector~$\boldsymbol{k}$. Together with the other linearized equations in Fourier space, we can show that the independent components of $\delta\boldsymbol{Q}_{\boldsymbol{k}}$ and the Fourier density perturbations $\delta\rho_E^{\boldsymbol{k}}$ and $\delta\rho_H^{\boldsymbol{k}}$ form a closed set of linear equations (as derived in  {Appendix~C.2}). In three spatial dimensions, this leads to a linear dynamical system of dimension seven. Two of its eigenvalues are identical to the eigenvalues of $\boldsymbol{\mathcal{S}}_k$, three are purely diffusive and equal to $-k^2$, and the remaining two are given by
	\begin{equation}
		{\lambda_I = -k^2+ \frac{\alpha\bar{\rho}_E(\Gamma+\eta\,k^2)}{3\left[(\Gamma+\eta\,k^2)+\Gamma\eta\left(\bar{\rho}_E+\bar{\rho}_H\right)\right]}.}
	\end{equation} This shows that the isotropic state is always unstable at low wavenumbers for extensile active stresses ($\alpha>0$). As shown in Fig.~\ref{fig:figures/figure_3}(a), the fastest growing mode is found to be $k=0$ if the activity parameter
	\begin{equation}
		\label{eqn:alpha_star_def}
		{\alpha<\alpha_\star = \frac{3\Gamma[1+\eta(\bar{\rho}_E+\bar{\rho}_H)]^2}{\eta^2\bar{\rho}_E(\bar{\rho}_E+\bar{\rho}_H)}}.
	\end{equation} If $\alpha>\alpha_\star$, we find that the fastest growing mode corresponds to a finite, nonzero wavenumber $k=k_0$, which is obtained by solving $\lambda_I'(k_0)=0$. An analysis of how the instability of the isotropic state depends on the spatial dimension is discussed in  {Appendix~C.3}, which also provides the linear instability analysis of other homogeneous base states in $\boldsymbol{Q}$, such as an uniaxial aligned state. This is closely related to the standard instability of active nematics, where active stresses destabilize ordered states, driving spontaneous flow and defect formation~\cite{AditiSimha2002, Marchetti2013, Saintillan2008}.
	\\
	\indent
	To investigate the role of active stresses beyond the linear regime, we numerically solve the system of equations (Table I) for different values of  $\alpha$. As before, the system is initially perturbed about the stationary and homogeneous state, with the density fluctuations having a vanishing mean; that is, the number of components remains unchanged and given by $N_{E,H} = \bar{\rho}_{E,H}V$. Furthermore, we choose initial data for the nematic order parameter such that its mean  is given by $V^{-1}\!\int\boldsymbol{Q}(\boldsymbol{r},0)\,\mathrm{d}^3\boldsymbol{r} = \frac{1}{3}\boldsymbol{I}$. Figure~\ref{fig:figures/figure_3}(e) presents snapshots of the heterochromatin density at long times for {$\alpha = 50$, $100$, and $150$}, with the corresponding dynamics shown in  {Videos~4, 5, and~6}. While for $\alpha=0$, the heterochromatin coarsens and forms a stable spherical droplet, for $\alpha\neq0$, droplets initially form and merge into larger droplets, undergoing continuous shape deformations. {At high activity levels,  the droplets are significantly deformed and stretched out, occasionally splitting into smaller droplets}. This highlights how active stresses introduce persistent fluctuations, preventing the emergence of stable, steady-state structures.  {On the other hand, euchromatin spans the entire domain and is only locally depleted within heterochromatin droplets (see SM, Figs.~S.2–S.5, for planar slices of $\rho_{E,H}$).}
	\\
	\indent
	As before, this can be quantified by computing the interface area $A$ of all heterochromatin droplets as function of time, normalized by $A_0=\sqrt[3]{36\pi }\,V_0^{2/3}$, with $V_0$ as the enclosed volume of all droplets. {As shown in Fig.~\ref{fig:figures/figure_3}(d), the ratio $A/A_0$ fluctuates around an average value greater than one, and this average increases with increasing activity.  {The coarsening dynamics, however, are non-monotonic in activity: at low activity (e.g., $\alpha=50$), active flows accelerate droplet encounters and mergers, so $A/A_0$ decays faster than in the inactive case ($\alpha=0$); at higher activity (e.g., $\alpha=100$), strong stresses continually deform and split droplets, regenerating interface, and $A/A_0$ instead saturates at an elevated level, remaining above both $\alpha=0$ and $\alpha=50$. Therefore, activity can both facilitate and hinder coarsening.}
	\\
	\indent
	The two-point correlation function of the solvent velocity $\boldsymbol{u}_S$ shows substantial dependence on $\alpha$, as shown in Fig.~\ref{fig:figures/figure_3}(b).
	The covariance $\textit{Cov}\left[\boldsymbol{u}_S,\boldsymbol{u}_S\right]$ is defined analogously to Eq.~(\ref{eqn:cov_rhoE_rho_H}), where the product between the velocity fields is interpreted as a vector dot product. {We define the correlation length as the radial separation at which the correlation function decays to $1/2$. This provides a consistent measure of the spatial range of correlations without assuming a specific functional form. The inset of Fig.~\ref{fig:figures/figure_3}(b) shows the correlation length $\xi_S$ of the solvent velocity, which initially increases sharply with increasing $\alpha$, before gradually saturating and exhibiting a slight decline at higher values of $\alpha$.}
	\\
	\indent
	While the correlation lengths of euchromatin and heterochromatin velocities vary with $\alpha$, a clearer and more interpretable trend emerges from the barycentric velocity, defined as 
	$\boldsymbol{u}_C = ({\rho_E\,\boldsymbol{u}_E + \rho_H \,\boldsymbol{u}_H})/{\rho_T}$. 
	{As shown in Fig.~\ref{fig:figures/figure_3}(c), the corresponding correlation length $\xi_C$ in $\boldsymbol{u}_C$ displays a pronounced increase with increasing $\alpha$, followed by a gradual saturation and a slight decline at larger values, as highlighted in the inset.} This behavior indicates that low to moderate activity levels enhance the spatial coherence of chromatin dynamics, facilitating collective motion over extended distances. 
	\\
	\indent
	To understand the euchromatin dynamics induced by active stresses, we analyze the nematic order parameter. The tensor field $\boldsymbol{Q}$ is characterized by its maximal nonnegative eigenvalue  $\lambda_\mathrm{max}$ which satisfies $1/3\le \lambda_\mathrm{max}\le1$. Assuming that $\lambda_\mathrm{max}$ is isolated, we define its associated unit eigenvector as the director field $\boldsymbol{n}$, and we introduce the scalar order parameter: ${q} = (3\lambda_\mathrm{max}-1)/2$. By these definitions, in the sharply aligned limit, we have that ${q}=1$ and $\boldsymbol{Q} = \boldsymbol{n}\boldsymbol{n}$, whereas ${q}=0$ in the isotropic state. In Fig.~\ref{fig:figures/figure_4}(b), we plot the scalar order parameter for the same configuration and simulation time as those in Fig.~\ref{fig:figures/figure_3}(e), revealing large, highly aligned regions, with ${q}$ close to $1$.  {Additional planar slices of $q$ are shown in SM, Figs.~S.6–S.9. The snapshots in Fig.~\ref{fig:figures/figure_4}(b) correspond to a statistically steady state, reached after an initial transient during which the volume-averaged scalar order parameter relaxes to a constant mean and fluctuates around it, see Fig.~S.12(a) in SM}. The size of the highly aligned regions can be quantified by computing the two-point spatial correlation in ${q}$ and determining its associated correlation length {$\xi_q$}, see Fig.~\ref{fig:figures/figure_4}(a). The results indicate that the correlation length {$\xi_q$ of the scalar order parameter} decreases as $\alpha$ increases. A similar trend is observed for the {correlation length $\xi_n$ of the director field}, which also decrease with increasing $\alpha$, see Fig.~\ref{fig:figures/figure_4}(c).  {We also find that the spatial correlation of $q$ and $\rho_H$ is short-range anticorrelated, with a correlation length that decreases with activity $\alpha$, as shown in SM, Fig.~12(b).}
	\\
	\indent
	To visualize the orientation of the director field within the highly aligned regions, as depicted in Fig.~\ref{fig:figures/figure_4}(b), we map each orientation on the unit sphere to a distinct color, ensuring that the antipodal points share the same color. This mapping, while not unique, is achieved by interpolating between three base colors--- {cyan, magenta, and yellow}---based on the squared components of the director field $\boldsymbol{n} = (n_x, n_y, n_z)$. Specifically, the fractions $n_x^2$, $n_y^2$, and $n_z^2$ determine the contributions of  {cyan, magenta, and yellow}, respectively, as illustrated in the legend in Fig.~\ref{fig:figures/figure_4}(d). This color scheme reveals coherent structures within the director field, highlighting that each highly aligned region is also found to be spatially oriented in the same direction. These are plotted in Fig.~\ref{fig:figures/figure_4}(d), along with the highly dense regions of heterochromatin, shown in dark gray (see legend). The snapshots correspond to the dynamics shown in  {Videos~7, 8, and 9}.  {To aid interpretation of the color map, planar slices of the director field $\boldsymbol{n}$ are shown alongside $q$ in SM, see Figs.~S.6–S.9, where $\boldsymbol{n}$ is projected onto the slice and overlaid as short rods whose length indicates the in-plane component.}

	\begin{figure*}[t!]\includegraphics[width=\textwidth]{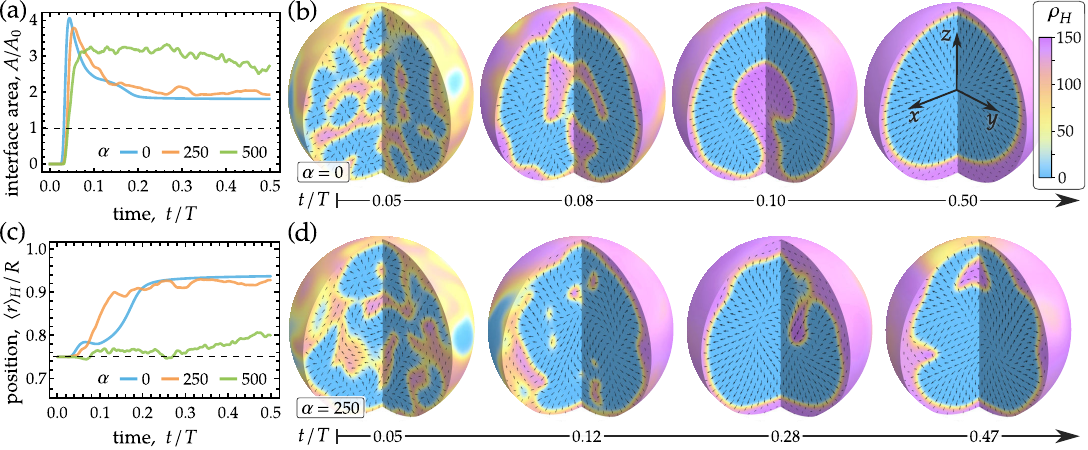}
		\caption{\label{fig:figures/figure_5} 
			(a) Total interface area $A$ of heterochromatic droplets in a spherical nucleus, rescaled by $A_0 =\sqrt[3]{36\pi }\,V^{2/3}_0$, where $V_0$ is the volume of all heterochromatic droplets at time $t/T$. Note that $A/A_0$ does not approach one, even in the absence of activity ($\alpha=0$). Herein, {$\eta=0.01$, $\bar{\rho}_C=100$, $\Gamma=25$, $\bar{\rho}_E =10$, and $\bar{\rho}_H =40$.} 
			(b) Snapshots of the heterochromatin density $\rho_H$ (see color bar) and its associated projected velocity $\boldsymbol{u}_H$ onto the surface of a spherical cut, with $\alpha=0$. As before, we observe coarsening of the initial droplets; however, at long times, the material is completely localized near the nuclear boundary, wetting the entire surface and forming a spherical shell. The shell thickness is determined by the initial amount of heterochromatin. 
			(c)~Time evolution of the averaged radial position of heterochromatin, $\langle r\rangle_H = N_H^{-1}\!\int r\hspace{1pt}\rho_H(r,\theta,\varphi)\,\mathrm{d}V$, rescaled by the nucleus radius~$R$. (d)~Snapshots of $\rho_H$ for nonzero activity strength ($\alpha={250}$), using the same color bar as in (b), and the corresponding projected velocity fields of heterochromatin  $\boldsymbol{u}_H$ onto the spherical cuts. 
		}
	\end{figure*}
	
	\section{Heterochromatin Wetting at Nuclear Boundary}
	
	So far, we have investigated the behavior of the system in an open, periodic domain, which enabled the study of linear instabilities around a homogeneous state and the analysis of the nonlinear regime through simulations. These revealed the formation of heterochromatin droplets and the emergence of large, highly aligned regions of euchromatin activity. We now aim to extend this study to a closed domain with an undeformable boundary, such as the nuclear envelope, which is assumed to be a sphere of radius $R$. In this configuration, the governing equations (see Table I) must be supplemented with boundary conditions at the nuclear boundary. By conservation of euchromatin and heterochromatin in the nucleus, no-flux boundary conditions for their densities are imposed:
	\begin{equation}
		\label{eqn:BC-no-flux}
		\boldsymbol{\hat{r}}\cdot\boldsymbol{\nabla}\rho_{E}\Big|_{|\boldsymbol{r}|=R}= 0,\quad\text{and}\quad \boldsymbol{\hat{r}}\cdot\boldsymbol{\nabla}\rho_H\Big|_{|\boldsymbol{r}|=R}= 0,
	\end{equation} where $\boldsymbol{\hat{r}}$ is the outward normal unit vector to the nuclear spherical surface. Moreover, we require a boundary condition on the nematic order parameter $\boldsymbol{Q}$ at the nuclear envelope. Assuming a conservation of molecular motors in the nucleus, we require a no-flux boundary condition:
	\begin{equation}
		\label{eqn:BC-no-flux-Q} \boldsymbol{\hat{r}}\cdot\boldsymbol{\nabla}\boldsymbol{Q}\\\Big|_{|\boldsymbol{r}|=R}= 0.
	\end{equation}
    
	Lastly, the momentum equations must be augmented with boundary conditions.  {Here, we impose} a no-slip condition on each fluid velocity:
	\begin{equation}
		\label{eqn:BC-no-slip}
		\boldsymbol{u}_S = 0,\quad\boldsymbol{u}_E = 0,\quad\text{and}\quad\boldsymbol{u}_H = 0, 
	\end{equation} on the nuclear surface ${|\boldsymbol{r}|=R}$.  {At the microscopic level, the nuclear envelope is reinforced by the lamin network, a meshwork of intermediate filaments bound to the inner membrane that interacts with chromatin. Here, we impose impermeable, no-slip boundary conditions as a coarse-grained description of interfacial friction, such that both chromatin fluids have zero velocity at the boundary. Although euchromatin and heterochromatin may differ in their molecular affinity for lamins~\cite{Belmont2017}, the present model does not account for such differences, and both phases are treated identically at the boundary.}
    \\
	\indent
	\begin{figure*}[t!]\includegraphics[width=\textwidth]{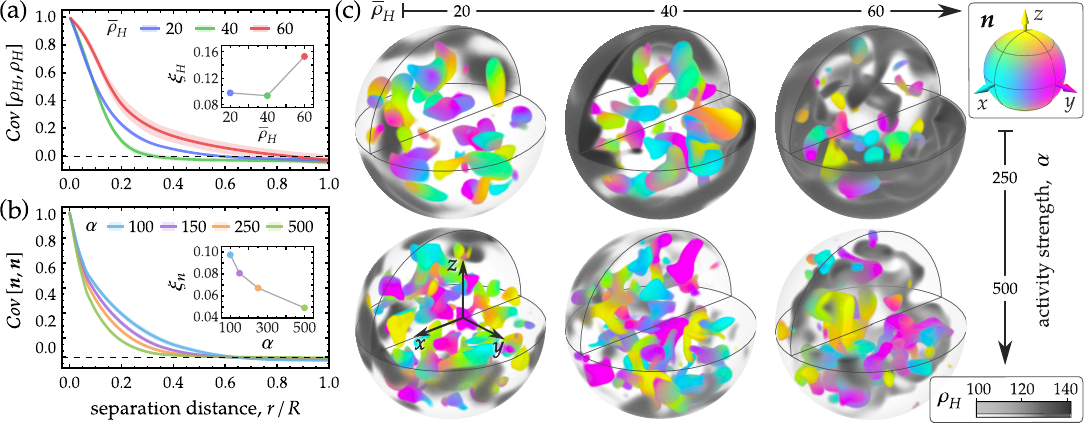}
		\caption{\label{fig:figures/figure_6} 
         {(a)~Two-point correlation of heterochromatin density fluctuations from the mean, and normalized by their variance, at different values of $\bar{\rho}_H$, with ${\bar\rho}_E=10$ and $\alpha=500$. This is computed for all points in the sphere of radius $R$ at a separation distance $r$. Hereinafter, we use {$\eta=0.01$,  $\bar{\rho}_C=100$, and $\Gamma=25$}. {Its associated correlation length $\xi_H$ (in units of $R$) is shown in the inset}. 
        (b)~Two-point correlation of director field fluctuations from the mean for different values of $\alpha$, at a separation distance $r/R$, using $\bar{\rho}_H = 40$ and the same parameters as in (a). The inset shows its correlation length $\xi_n$ (in units of $R$).} 
        (c)~Three-dimensional plots of the  director field $\boldsymbol{n}$ by varying $\alpha$ and density $\bar{\rho}_H$, at fixed $\bar{\rho}_E$, and using the color scheme shown in the inset---the same as in Fig.~\ref{fig:figures/figure_4}(d). Here, only the points with ${q}\geq0.96$ are shown, while the rest are transparent. Dark gray shows the high density values of the heterochromatin; see legend for color and opacity levels.
		}
	\end{figure*}
	By using the boundary conditions in Eq.~(\ref{eqn:BC-no-flux}),  (\ref{eqn:BC-no-slip}), and (\ref{eqn:BC-no-flux-Q}), we solve numerically the governing equations. Here, we make use of a fast spectral method (Dedalus) with domain discretization in spherical coordinates and an implicit-explicit fourth order Runge-Kutta scheme for time integration \cite{Dedalus}. As before, we start from a stationary state (with $\boldsymbol{u}_{S,E,H} =0$ everywhere in the nucleus) and initially apply a perturbation in the chromatin densities about the homogeneous state ($\bar{\rho}_E$ and $\bar{\rho}_H$), and a perturbation in the nematic order parameter about the isotropic state. Note that the initial perturbations are chosen such that the total number of euchromatin and heterochromatin components remains unchanged; namely, we have $N_{E,H} = \bar{\rho}_{E,H} V_N = \iiint\!\rho_{E,H}\,\mathrm{d}V$, where $V_N = \frac{4\pi}{3}R^3$ is the nuclear volume, and the integral is taken over the whole nucleus. Similarly, the initial perturbation in the nematic order parameter is chosen such that its mean value is given by $V_N^{-1}\iiint\boldsymbol{Q}\,\mathrm{d}V = \frac{1}{3}\boldsymbol{I}$.
	\\
	\indent   
    We begin by investigating the system in the absence of euchromatin activity ($\alpha = 0$). The homogeneous densities \(\bar{\rho}_E\) and \(\bar{\rho}_H\) are chosen, along with parameters \(\eta\), \(\bar{\rho}_C\), and \(\Gamma\), such that the system becomes linearly unstable, see Eq.~(\ref{eqn:condition-instability}). This drives the formation of heterochromatin droplets, which rapidly coalesce into larger structures. Unlike the behavior observed in an open, periodic system, where a single large droplet typically forms, the heterochromatin in a closed domain redistributes along the nuclear boundary.  {To assess this difference more directly, we compared spherical and periodic domains under identical parameters (see SM, Fig.~S.11). Both show droplets of comparable size during the early growth phase, consistent with the linear instability, while confinement in the sphere affects only the later coarsening, leading to surface wetting.} Snapshots of this are depicted in Fig.~\ref{fig:figures/figure_5}(b), which are taken from the numerical simulation shown in  {Video~10} (with $\bar{\rho}_H = 40$). The extent of surface wetting and the thickness of the heterochromatin layer depend on the total heterochromatin content, quantified by $N_H\!=\!\bar{\rho}_H V_N$, and the  {saturation} density \(\rho_H^\mathrm{max}\) achieved at long times {(see  {Video~11~and~12}, where $\bar{\rho}_H = 10$~and~$20$, respectively)}. By assuming that heterochromatin uniformly covers the entire nuclear surface and that the interface is sharp, the thickness of the spherical heterochromatin layer can be analytically approximated: $\delta r_H= R\left(1 - \sqrt[3]{1 -  {\bar{\rho}_H}/{\rho_H^\mathrm{max}}}\right)$.  {We find that the saturation density $\rho_H^{\max}$ increases with $\bar{\rho}_C$. Varying $\bar{\rho}_C$ in simulations (see SM, Fig.~S.10) does not affect the peripheral localization of heterochromatin at long times but changes its shell thickness, which decreases with increasing $\bar{\rho}_C$, consistent with the analytical estimate for~$\delta r_H$.}    
    \begin{figure*}[t!]\includegraphics[width=\textwidth]{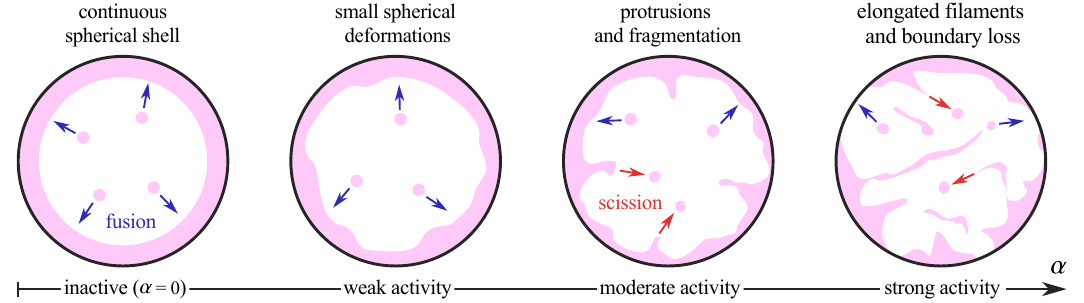}
    \caption{\label{fig:figure_diagram} 
     {Morphological regimes of heterochromatin in a spherical nucleus as a function of activity strength $\alpha$ (other parameters fixed), under the no-slip conditions. At $\alpha=0$, heterochromatin forms a continuous peripheral shell. Small $\alpha$ yields smooth, small-amplitude undulations of the interface. Moderate $\alpha$ produces larger deformations, with protrusions undergoing scission into smaller droplets. Large $\alpha$ generates highly dynamic morphologies with elongated filaments, stretched droplets, as well as partial boundary loss. Heterochromatin is shown in magenta. Blue and red arrows indicate fusion and scission events.}
    }\end{figure*} 
	\\
	\indent
	As before, the surface of the interface can be identified by its half-maximum density points, $\rho_H^\mathrm{max}/2$. Using surface triangulation, we compute the net area $A$, as shown in Fig.~\ref{fig:figures/figure_5}(a), normalized by the effective spherical area $A_0=\sqrt[3]{36\pi}\,V_0^{2/3}$, where $V_0$ is the total enclosed volume. Contrary to the case in an open, periodic domain, the ratio $A/A_0$ does not converge to one. Instead, we find that $A/A_0$ approaches $({\rho_H^\mathrm{max}}/{\bar{\rho}_H}-1)^{2/3}$ 
	at long times when heterochromatin wets entirely the nuclear surface boundary. This limit follows directly from the exact expression derived for the heterochromatin thickness $\delta r_H$.
	\\
	\indent
	Fig.~\ref{fig:figures/figure_5}(c) depicts the averaged radial position of heterochromatin, $\langle r\rangle_H = N_H^{-1}\!\int r\,\rho_H(r,\theta,\varphi)\,\mathrm{d}V$, where $r$ is the radial position within the nucleus, $\theta$ and $\phi$ are the inclination and azimuthal angles, and the volume element $\mathrm{d}V = r^2\sin\theta\mathrm{d}\theta\mathrm{d}\varphi$. The figure shows  directly the mean radial position of the interface, with heterochromatin wetting entirely the nuclear surface at long times.
	\\
	\indent
     {For comparison, we additionally solved the system with no–penetration and stress–free tangential conditions on both chromatin phases, while retaining no-slip for the solvent. Under this choice, heterochromatin coarsens into a single, compact, spherical-cap–like domain that merely contacts the spherical boundary rather than forming a peripheral shell (Video~13). This indicates that some form of frictional tangential boundary condition is necessary for peripheral redistribution of heterochromatin, with no-slip providing the minimal mechanism.}
    \\
	\indent
	We now solve the governing equations for extensile activity ($\alpha > 0$) and find that heterochromatin no longer uniformly wets the nuclear surface at long times. Instead, its interface undergoes continuous deformations, forming protrusions that often break away and rapidly remerge with the wetted region. The dynamics of this behavior are shown in  {Videos~14 and 15}, corresponding to $\alpha = 250$ and $500$, respectively. Snapshots illustrating this, taken from  {Video~14} ($\alpha = 250$), are shown in Fig.~\ref{fig:figures/figure_5}(d). 
    \\
	\indent
	The interface deformations result in fluctuations of the area ratio $A/A_0$ and the average radial position of heterochromatin, as shown in Fig.~\ref{fig:figures/figure_5}(a) and (c), respectively, highlighting how activity disrupts the distribution of heterochromatin along the boundary. The heterochromatin dynamics can be further quantified by measuring the two-point spatial correlation of the heterochromatin density, $\textit{Cov}\left[\rho_H, \rho_H\right]$, defined similarly to Eq.~(\ref{eqn:cov_rhoE_rho_H}), with the integrals now taken over the entire nuclear volume.  {Fig.~\ref{fig:figures/figure_6}(a)} shows this correlation averaged over various time snapshots, excluding the early dynamics ($t/T\geq0.1$), where we vary the net amount of heterochromatin at fixed values of $\alpha$ and $\bar{\rho}_E$. {The inset of  {Fig.~\ref{fig:figures/figure_6}(a)} shows that the correlation length $\xi_H$ remains relatively short at intermediate heterochromatin densities but increases sharply at higher densities under high activity. This suggests that active processes, combined with crowding, lead to a more uniform distribution of heterochromatin.}
    \\
	\indent
	The time-averaged correlation associated with the director field $\boldsymbol{n}$ is shown in Fig.~\ref{fig:figures/figure_6}(b), which reveals that its correlation length $\xi_n$ decreases as we increase the activity $\alpha$ (see inset), at fixed density $\bar{\rho}_H$. Similar to the open system, the closed system exhibits large, highly aligned regions (with ${q}$ near $1$) that share closely the same director field orientation $\boldsymbol{n}$. Fig.~\ref{fig:figures/figure_6}(c) illustrates these coherent structures, showing the highly aligned regions alongside areas of high heterochromatin density {by varying activity levels, $\alpha$, and heterochromatin content, $\bar{\rho}_H$, at fixed euchromatin density $\bar{\rho}_E$.} The corresponding dynamics for $\alpha = 250$ are presented in  {Videos~16, 17, and~18}, which correspond to $\bar{\rho}_H = 20$, $40$, and $60$, respectively  {(equatorial-plane slices of $q$ and $\boldsymbol{n}$ corresponding to these configurations are shown in SM, Figs.~S.13–S.14). Snapshots in Fig.~\ref{fig:figures/figure_6}(c) are taken from a steady regime, where the volume average of $q$ over the sphere fluctuates around a constant mean; see SM, Fig.~S.12(c)}. These results reveal a complex interplay between $\alpha$ and $\bar{\rho}_H$ in shaping the heterochromatin profile. At low density $\bar{\rho}_H$, heterochromatin forms disconnected patches that wet the nuclear surface, which are moved along the boundary by the euchromatin activity. As $\bar{\rho}_H$ increases, heterochromatin becomes sufficient to coat entirely the nuclear surface, although euchromatin activity may intermittently displace it from the boundary. At even higher $\bar{\rho}_H$, the heterochromatin layer thickens, compressing the euchromatin into a smaller volume. This increased euchromatin density amplifies the active stress, further deforming the heterochromatin interface. Thus, even without explicitly increasing $\alpha$, heterochromatin can indirectly enhance the euchromatin activity by reducing its available volume. In extreme cases of both high heterochromatin content and large activity, euchromatin may displace heterochromatin entirely from the nuclear boundary.  {The limited overlap between heterochromatin and highly aligned regions in Fig.~\ref{fig:figures/figure_6}(c) suggests an anticorrelation between $q$ and $\rho_H$ (that is, regions of high $q$ coincide with low heterochromatin density). As shown in SM Fig.~S.12(d), the spatial correlation between $q$ and $\rho_H$ exhibits short-range anticorrelation and a correlation length that decreases with $\alpha$, reflecting stronger active stirring and fragmentation of heterochromatin at higher activity.}

	\vspace{-6pt}\section{Discussion}

	 {Our continuum framework captures how active, contractile mechanics drive the formation of chromatin compartments, leading to the demixing of euchromatin and heterochromatin and their ensuing morphological changes}. We show that heterochromatin can spontaneously form droplets through a mechanically driven phase separation governed by frictional, viscous, and contractile forces. This demixing is primarily controlled by contractile stresses arising from heterochromatin cross-linking proteins, leading to the formation of nascent droplets that subsequently coalesce into larger ones. This provides a physical mechanism for chromatin phase separation, where heterochromatin-rich regions segregate from transcriptionally active euchromatin-rich domains.
    \\
	\indent
    In an open or periodic system, such coarsening typically yields a single large heterochromatin droplet.  {In a confined spherical nucleus, by contrast, heterochromatin relocates to the nuclear periphery, wetting the boundary and forming a shell-like layer; this matches observations across many cell types in which heterochromatin is enriched at the periphery \cite{Bizhanova2021}. While earlier polymer models reproduced this enrichment by assuming direct binding to the nuclear lamina \cite{Mirny2019, Saintillan2022}, it has also been shown that even a nonspecific envelope attraction can favor peripheral heterochromatin owing to its higher compaction and greater areal density of binding elements~\cite{Spakowitz2020}. In our framework, both chromatin phases satisfy no-penetration and tangential no-slip boundary conditions. Interestingly, despite this non-preferential treatment, only heterochromatin accumulates at the periphery, consistent with~\cite{Spakowitz2020}. In contrast, if the tangential condition for both phases is replaced by a stress-free one, heterochromatin instead coarsens into a compact, spherical-cap–like domain that merely contacts the envelope (see Video~13).
    Thus, no-slip provides a minimal form of tangential anchoring, showing that some anchoring is required to sustain a peripheral shell of heterochromatin. Experimental studies \cite{Belmont2017, Solovei2013, Kind2013} show that heterochromatin generally has a higher affinity for the nuclear lamina. Such preferential binding could, in principle, be incorporated into our framework by adding a phase-specific surface-anchoring term to the tangential momentum balance at the boundary. This term would enter as an interfacial traction acting on the chromatin phase, with a strength proportional to the product of the lamin concentration and the boundary density of that phase. Depending on the relative magnitudes of these anchoring strengths, one might expect stronger heterochromatin anchoring to promote a more stable peripheral shell, whereas comparable euchromatin anchoring could lead to competition for boundary occupancy and reduced heterochromatin enrichment.}
    \\
	\indent
	Introducing euchromatin activity disrupts this peripheral localization,  { driving a sequence of distinct morphological regimes, as depicted in Fig.~\ref{fig:figure_diagram}. For small euchromatic activity, the peripheral heterochromatin shell develops smooth, small-amplitude undulations. At moderate activity, these deformations grow into protrusions that can undergo scission into smaller droplets. At large activity, the heterochromatic morphology becomes highly dynamic, with elongated filaments and stretched droplets undergoing frequent fusion and breakup, and regions of partial boundary loss. This phenomenology is consistent with cycling cells, where a peripheral heterochromatin layer coexists with interior droplets~\cite{Spector2010}. Nevertheless, the nuclear interior also contains other large subnuclear bodies (e.g., nucleoli, speckles) that interact with chromatin and can serve as internal anchoring sites \cite{Bizhanova2021}; their inclusion in our model could produce more realistic spatial patterns and further promote interior droplets.} 
    \\
	\indent
     {
    The progression shown in Fig.~\ref{fig:figure_diagram} arises from the competition between three mechanical effects: (i) heterochromatin contractility, which compacts and smooths the peripheral shell; (ii) active euchromatin stresses, which drive flows that destabilize the interface; and (iii) the presence of a confining boundary, which biases heterochromatin toward the nuclear periphery. Whether the heterochromatic shell remains stable, deforms, or fragments, depends on the relative magnitude of active stresses (set by $\alpha$) compared with response stresses from contractility and the effects of confinement. Interface undulations emerge when these contributions are comparable, while fragmentation or filament formation occurs when active stresses are larger in scale. Viewed in reverse, decreasing activity strengthens heterochromatin compartmentalization, which is consistent with ATP-depletion experiments that lead to chromatin condensation \cite{Zidovska2013}\footnote{Experiments using transcriptional inhibition alone typically show chromatin decondensation, which arises from activation of the DNA damage response rather than loss of transcription.}. If instead, the amount of heterochromatin is increased at fixed $\alpha$, as in Fig.~\ref{fig:figures/figure_6}(c), the shell thickens as heterochromatin occupies more nuclear volume. This confines euchromatin, raising its density and amplifying the active stress, even though $\alpha$ is unchanged; this destabilizes and remodels the heterochromatin much like raising $\alpha$. We note that all of these morphological changes are not driven by an explicit interfacial tension; rather, in our model, an effective surface tension emerges self-consistently from contractility and from stress and velocity gradients across the interface, and therefore {surface tension} is not an independent parameter.
    }    
    \\
	\indent
     {Another major consequence of euchromatin activity in our model is the emergence of long-range coherence in chromatin motion: as activity increases, the correlation length of the barycentric chromatin velocity grows to micron scales, comparable to the nuclear size. The main driver of this behavior is the extensile stress associated with euchromatin, which self-organizes into large coherent domains spanning extended regions of the nucleus.  Through hydrodynamic coupling, these domains act as coherent force densities that drive nucleoplasmic flows and entrain both chromatin phases.
    Accordingly, the solvent velocity shows an activity-dependent increase in its correlation length. These predictions are consistent with live-cell measurements, which reveal ATP-dependent chromatin motions over micron scales \cite{Zidovska2013} and show that, in low-compaction chromatin, a transcriptionally active locus moves coherently with the nearby chromatin over micron scales~\cite{Zidovska2024}. Such long-range coherence is difficult to explain by purely elastic stresses transmitted through the polymer network, which tend to weaken as the network is diluted. By contrast, experiments reveal strong correlations in active, low-compaction regions, consistent with solvent-driven hydrodynamics. Our results support this interpretation and are corroborated by prior active-polymer hydrodynamic models~\cite{Saintillan2018, Saintillan2022}.}
	\\
	\indent
	Previous continuum models that described chromatin as a two-fluid system, accounted for chromatin and nucleoplasm as two separate phases \cite{Bruinsma2014, Eshghi2023, Grosberg2023}. Our current approach distinguishes euchromatin and heterochromatin as two distinct compressible fluids, which interact with an incompressible nucleoplasm via friction. This distinction allows us to capture the separate mechanical roles of heterochromatin condensation and euchromatin-driven extensile stresses. In contrast to free-energy-based models \cite{Mirny2019, Safran2021, Safran2023, Potoyan2024}, which describe phase separation in equilibrium or near-equilibrium settings, our mechanical approach is based on mass and momentum conservation laws.  {This framework naturally accommodates both passive and active stresses, which enter the equations of motion for solvent and chromatin. Their coupling drives chromatin density variations and shapes nuclear organization. We note that the model focuses on the fluid-like behavior of chromatin observed at timescales longer than its elastic stress–relaxation time \cite{Eshghi2021}, neglecting the corresponding elastic response. Viscoelasticity, however, can introduce features absent from a purely fluid description, including stress storage and release, transient recoil, and frequency-dependent responses that alter instability thresholds. Such effects could slow the coarsening of heterochromatin droplets, stabilize intermediate morphologies, and maintain non-spherical shapes. These represent a natural extension for future studies.} 
    \\
    \indent 
     {Our coarse-grained formulation also omits molecular specificity (e.g., HP1 binding/unbinding, chromatin modifiers, ATP-dependent remodelers). Such factors might be effectively treated as regulatory concentration fields with their own advection–diffusion–reaction dynamics: unbound species advect with the solvent velocity $\boldsymbol{u}_S$, while chromatin-bound species advect with $\boldsymbol{u}_E$ or $\boldsymbol{u}_H$ and exchange with the unbound pool via binding/unbinding kinetics. These fields can modulate parameters such as the contractility parameter $\bar{\rho}_C$ or the activity strength $\alpha$, or drive interconversion between $\rho_E$ and $\rho_H$. Lamina tethering can be modeled similarly by a boundary-bound field with surface advection–diffusion–reaction, coupled to the bulk via phase-specific anchoring terms in the tangential momentum balance. Such extensions are compatible with our framework and would link biochemical regulation to the mechanical processes described here.}
	\\
	\indent
  	Taken together, our results  {show} how active processes, mechanical forces, and geometric constraints can collectively organize chromatin at the nuclear scale.  {These findings lead to testable predictions for how activity and confinement influence chromatin dynamics and morphologies, which can be probed experimentally in live cells. On the theoretical side, the framework could be extended to incorporate the viscoelasticity of chromatin, explicit molecular interactions, and further explorations of chromatin–lamina coupling. Such efforts will advance our understanding of how physical and biochemical mechanisms together regulate genome organization in the nucleus.}
    
    \begin{acknowledgments}
		{The authors acknowledge funding from the National Science Foundation Grants No.\ CMMI-1762506 and No. DMS-2153432 (A.\,Z.~and M.\,J.\,S.), No.\ CAREER PHY-1554880 and No.\ PHY-2210541 (A.\,Z.), and No.\ CMMI-1762566 and No.\  DMS-2153520 (D.\,S.). The authors also thank A.\ Lamson, S.\ Weady, and B.\ Chakrabarti for stimulating discussions and helpful feedback.  {Computations were carried out at the Flatiron Institute’s Scientific Computing Core.}
        }
	\end{acknowledgments}
	
	
	\appendix
	
	\section*{Appendix}\appendix
	
	\section{ {Dimensionless groups}}	
	
	By rescaling time and space using a characteristic time scale $T$ and a length scale $L$, respectively, the continuity equations can be rescaled, yielding a single dimensionless parameter: $d_c T/L^2$. Thus, by choosing that 
	\begin{equation}
		\label{eqn:time-scale}
		T = L^2/d_c
	\end{equation} the dimensionless continuity equations can be rewritten as in Table I. Note that we also rescaled the local densities by a characteristic density-scale $\rho_0$.
	
	Moreover, rescaling each of the momentum equations by {a force-density scale $f_0$}, we find the following dimensionless friction coefficient 
	\begin{equation}
		{\Gamma = \frac{\Gamma_0\rho_0 L}{f_0 T}}.
	\end{equation} From the momentum equation of the solvent, we get two additional dimensionless parameters: a rescaled activity 
	\begin{equation}
		{\alpha = \frac{\alpha_0\rho_0}{f_0 L}},
	\end{equation} and a rescaled solvent viscosity {$\eta_s/(f_0 L T)$}. Setting the latter to unity, we find the following {force-density scale}
	\begin{equation}
		\label{eqn:stress-scale}
		{f_0 = \frac{\eta_s}{L T} = \frac{\eta_s d_c}{L^3}},
	\end{equation} where the last equality follows from Eq.~(\ref{eqn:time-scale}). Also, from the momentum equations of euchromatin and heterochromatin, we find the following dimensionless parameters: first, a rescaled compressibility {$B\rho_0^2/(f_0 L)$}, which is set to unity and gives the following characteristic density 
	\begin{equation}
		\label{eqn:density-scale}
		{\rho_0 = \sqrt{\frac{f_0 L}{B}}= \frac{\sqrt{\eta_s d_c}}{L\sqrt{B}}};
	\end{equation} second, a rescaled chromatin viscosity coefficient
	\begin{equation}
		{\eta = \frac{\rho_0\eta_c}{f_0 L T} = \frac{\eta_c \sqrt{d_c}}{L\sqrt{\eta_s B}}},
	\end{equation} where the last equality follows from Eq.~(\ref{eqn:stress-scale}) and (\ref{eqn:density-scale}); and lastly a dimensionless cross-linking density
	\begin{equation}
		{\bar{\rho}_C = \frac{B{\rho}_C\rho_0}{f_0 L} =\frac{\rho_C}{\rho_0}= \frac{\rho_C L\sqrt{B}}{\sqrt{\eta_s d_c}}},
	\end{equation} where last two equalities follow from Eq.~(\ref{eqn:stress-scale}) and (\ref{eqn:density-scale}). The solvent pressure is rescaled as: {$p = L^2 p_S/(\eta_s d_c)$}.   
    Thus, by means of Eqs.~(\ref{eqn:time-scale}), (\ref{eqn:stress-scale}) and (\ref{eqn:density-scale}), the dimensionless friction coefficient $\Gamma$ can be written as
	\begin{equation}
		\label{eqn:Gamma-def}
		{\Gamma = \frac{\Gamma_0 L\sqrt{d_c}}{\sqrt{\eta_s B}}},
	\end{equation} whereas the dimensionless activity strength $\alpha$ reads
	\begin{equation}
		\label{eqn:Alpha-def}
		{\alpha = \frac{\alpha_0 L}{\sqrt{\eta_s d_c B}}}.
	\end{equation} 
    
    Note that a natural length-scale $L$ could be defined by setting either {$\Gamma$, $\alpha$ or $\bar{\rho}_C$ to one}; however, {here we choose $L$ to be given by the radius of the cell nucleus $R$}. Table I shows the continuity and momentum equations, and the dynamical equation of the nematic order parameter $\boldsymbol{Q}$, rescaled by their characteristic scales ($T$, $L$, and $\rho_0$), in terms of the dimensionless parameters $\alpha$, $\eta$, $\bar{\rho}_c$ and $\Gamma$.

	\section{ {Parameter estimation}}
    
     {Nuclear sizes vary across higher eukaryotes: in human interphase cells, equivalent-sphere radii are found to be in the range of $4$--$7\,\mu\mathrm{m}$~\cite{Milo2015}, while some large embryonic cells can exceed $10\,\mu\mathrm{m}$~\cite{Heald2010}. Here, we use radius $R=5\,\mu\mathrm{m}$ as a conservative choice.}

     {Chromatin tracer diffusivity measured by fluorescence correlation spectroscopy (FCS) and live-cell tracking lie in the range $10^{-3}$--$10^{-2}\,\mu\mathrm{m}^2/\mathrm{s}$~\cite{Shaban2020, Barth2020, Daugird2024}. This corresponds to a time scale $T = R^2/d_c$ spanning $0.7$--$14\,\mathrm{hrs}$. We take the geometric mean $d_c = 3.2\times10^{-3}\mu\mathrm{m}^2/\mathrm{s}$ as a representative value, which gives $T \approx 3.1\,\mathrm{hrs}$. }
    
     {Nucleoplasmic viscosity $\eta_s$ has been estimated with inert $2$--$10\,\mathrm{nm}$ probes in FCS and fluorescence recovery after photobleaching (FRAP), with reported values in the range $3$--$10\,\mathrm{mPa}\cdot\mathrm{s}$~\cite{Siebrasse2008,Liang2009,Erdel2015,Hertzog2023}. We take the geometric mean $\eta_s = 5.5\,\mathrm{mPa}\cdot\mathrm{s}$ as a representative value.}

     {Euchromatin density can be estimated (in terms of nucleosomes per unit volume) from the genome size, which is $6.4\,\mathrm{Gbp}$ for human diploid~\cite{Milo2015}. By assuming $0.5$--$0.9$ of the genome is euchromatin, and taking $200\,\mathrm{bp}$ per nucleosome, we find the euchromatin density to be in the range of $1.1\times10^{4}$\,--\,$5.5\times10^{4}\,\mu\mathrm{m}^{-3}$. Since densities in the model are expressed relative to the characteristic scale $\rho_{0}$, in simulations we set the initial homogeneous euchromatin density to $\bar\rho_{E}=10$ (in dimensionless units). This choice fixes $\rho_{0}$, and we use the geometric mean as a representative value $\rho_{0}\approx2.5\times10^{3}\,\mu\mathrm{m}^{-3}$. This rescaling ensures that the dimensionless densities in the simulations correspond to physiological values. By construction $\rho_{0}$ is defined such that $B\rho_{0}^{2}R^{2}/(\eta_{s}d_{c})=1$. Hence, fixing $\rho_{0}$ from the euchromatin density implicitly sets $B$, which lies in the range $10^{-12}$\,--\,$3\times10^{-9}\,\mathrm{mPa}\cdot\mu\mathrm{m}^{6}$. Here, we use the geometric mean, $B \approx 10^{-10}\,\mathrm{mPa}\cdot\mu\mathrm{m}^{6}$, as a working estimate. We emphasize that this is not a microscopic prediction, but an effective parameter determined self-consistently from the choice of density scale.}

     {Another parameter we fix in the model is the dimensionless contractility $\bar{\rho}_C = 100$. Since $\rho_C = \bar{\rho}_C\,\rho_0$, this corresponds to $\rho_C \approx 2.5 \times 10^{5}\,\mu\mathrm{m}^{-3}$. The corresponding isotropic crosslinking pressure is given by $\tilde{p}_C = B \rho_C \rho_H$, which for $\rho_H = 150$ (highly condensed heterochromatin in our simulations) yields $\tilde{p}_C \approx 7.4\,\mathrm{mPa}$. This can be compared with the steric pressures $\tilde{p}_{E,H} = -\tfrac{1}{2}B \rho_{E,H}\,\rho_T$, where $\rho_T = \rho_E + \rho_H$ is the total density. In pure euchromatin ($\rho_H = 0$), the crosslinking pressure $\tilde{p}_C$ vanishes identically, and the steric term is a small baseline pressure, $\tilde{p}_E \approx -0.1,\mathrm{mPa}$.  In heterochromatin-rich regions, the steric term $\tilde{p}_H \approx -5.7,\mathrm{mPa}$, which competes with the crosslinking pressure, and their contributions are of the same order (a few mPa).  }

     {Microscopically, the crosslinking pressure arises from HP1 dimers bridging pairs of H3K9me3 nucleosomes~\cite{Machida2018}. We assume that a bridge exerts a force $f_{\mathrm{br}}$ over a lever $\ell_{\mathrm{br}}$, corresponding to a force dipole $f_{\mathrm{br}}\,\ell_{\mathrm{br}}$, which results in an isotropic pressure $\tilde{p}_C = \tfrac{1}{3}c_{\mathrm{br}} f_{\mathrm{br}}\,\ell_{\mathrm{br}}$ by averaging over all dipole orientations, where $c_{\mathrm{br}}$ is the density of bridges. We take  $\ell_{\mathrm{br}} = 10$--$20$\,nm as the spacing of adjacent nucleosomes~\cite{Machida2018}. Experiments on HP1--DNA condensates with optical tweezers reveal force–extension curves with characteristic responses of several to tens of pN, depending on DNA length and condensate size~\cite{Keenen2021}; we take conservatively the single-bridge force to lie in the range of $f_{\mathrm{br}} = 1$--$5\,\mathrm{pN}$. Based on quantitative proteomics \cite{Itzhak2016}, the total HP1 nuclear concentration is estimated to be about $c_{\mathrm{HP1}} = 1\text{--}3\,\mu\text{M}$ (or $600$--$1800$ molecules$/\mu\mathrm{m}^3$). FRAP experiments shows that only a fraction $\phi_{\mathrm{bound}} = 0.1\text{--}0.3$ of this pool is bound to chromatin~\cite{Muller2009, Schmiedeberg2004}. Hence, the density of bridges can be expressed as $c_{\mathrm{br}} = c_{\mathrm{HP1}}\,\phi_{\mathrm{bound}}\,\phi_{\mathrm{br}}$, where $\phi_{\mathrm{br}}$ is the probability that a bound dimer simultaneously engages two nucleosomes. This probability is not directly known; here, we take $\phi_{\mathrm{br}} = 0.1$--$0.5$ as a plausible range, since bound dimers may lack neighbors or exchange too rapidly to form bridges. This yields a bridge density of about $6$--$270\,\mu\mathrm{m}^{-3}$, corresponding to a crosslinking pressure in the range of $\tilde{p}_C = 0.1$--$9\,\mathrm{mPa}$, which is consistent with the effective value of $\sim \!7\,\mathrm{mPa}$ implied by our nondimensionalization. }

     {The solvent viscosity $\eta_s$ and friction coefficient $\Gamma_0$ define a solvent screening length $\ell_s = [\eta_s/(\Gamma_0 \rho_T)]^{1/2}$, which characterizes how far hydrodynamic disturbances propagate before being damped. Since $\ell_s$ decreases with increasing chromatin density, propagation is shorter-ranged in heterochromatin-rich regions than in euchromatin. As $\ell_s$ is not directly measurable, we take it to be proportional to structural scales of the interchromatin space, as expected from porous-medium hydrodynamics. Single-particle tracking of inert probes reveals confinement and caging on the order of $100\,\mathrm{nm}$, and electron microscopy together with super-resolution imaging show interchromatin channels of about $100$--$300\,\mathrm{nm}$~\cite{Tseng2004,Goerisch2004,Hameed2012,Ou2017,Miron2020}. These length scales place the dimensionless friction coefficient $\Gamma$ in the range $20$--$500$ (that is, $\Gamma_0 = 1\text{--}40\,\mathrm{mPa}\cdot\mathrm{s}\cdot\mathrm{nm}$). In simulations, we fix $\Gamma=25$, which corresponds to a screening length $\ell_s \approx 0.3\,\mu\mathrm{m}$ (being near the upper end of the range). Alongside the solvent screening length, we define a chromatin screening length $\ell_{c}=\sqrt{\eta_c/\Gamma_0}=R\sqrt{\eta/\Gamma}$, which characterizes the decay length of chromatin momentum under solvent drag. We assume it is of the same order but smaller than $\ell_s$; here, we set to $\ell_{c}\approx0.1~\mu\mathrm{m}$, which fixes the dimensionless chromatin viscosity coefficient at $\eta\approx0.01$. The chromatin viscosity scales linearly with phase density, giving an euchromatin viscosity $\eta_E=\eta_c\rho_E\approx0.1\,\eta_s$ (in pure euchromatin, $\rho_E=10$) and a heterochromatin viscosity $\eta_H=\eta_c\rho_H\approx1.5\,\eta_s$ (in the dense regions of heterochromatin, $\rho_H =150$). }

    \begin{table}
    \begin{ruledtabular}\begin{tabular}{ll}
        &\\[-8pt]
        Nucleus radius & 
        \; $R = 4$--$7~\mu\mathrm{m}$~\cite{Milo2015}\\[4pt]
        Chromatin diffusivity & 
        \; $d_c = 10^{-3}$--$10^{-2}\,\mu\mathrm{m}^2/\mathrm{s}$~\cite{Shaban2020, Barth2020, Daugird2024}\\[4pt]
        Diffusive time-scale & 
        \; $T = 1$--$14\,\mathrm{hrs}$\\[4pt]
        Nucleoplasm viscosity & 
        \; $\eta_s = 3$--$10\,\mathrm{mPa}\cdot\mathrm{s}$~\cite{Siebrasse2008,Liang2009,Erdel2015,Hertzog2023}\\[4pt]
        Characteristic density \qquad &
        \; $\rho_{0}= 1.1$--$5.5\times10^{3}\,\mu\mathrm{m}^{-3}$\\[4pt]
        Compressibility coefficient &
        \; $B = 10^{-12}$--$10^{-9}\,\mathrm{mPa}\cdot\mu\mathrm{m}^{6}$\\[4pt]
        Contractility parameter \qquad &
        \; $\rho_{C}= 1.1$--$5.5\times10^{5}\,\mu\mathrm{m}^{-3}$\\[4pt]
        Friction coefficient &
        \; $\Gamma_0 = 1$--$40\,\mathrm{mPa}\cdot\mathrm{s}\cdot\mathrm{nm}$\\[4pt]
        Chromatin viscosity &
        \; $\eta_c = 5$--$90\times10^{-6}\,\mathrm{mPa}\cdot\mathrm{s}\cdot\mu\mathrm{m}^{3}$\\[4pt]
        Activity strength &
        \; $\alpha_0 = 10^{-4}$--$10^{-1}\,\mathrm{pN}\cdot\mathrm{nm}$\\[4pt]
    \end{tabular}\end{ruledtabular}
    \caption{\label{tab:parameters}  {Dimensional parameters and their estimated physiological ranges. Values are drawn from experimental measurements where available and otherwise estimated from theoretical considerations and scaling arguments.}}
    \end{table}
    
     {To estimate the activity strength $\alpha_0$ of the active stress $\boldsymbol{\sigma}_A=-\alpha_0\rho_E \boldsymbol{Q}$, we make the microscopic identification $\alpha_0\rho_E=n_a f_a\ell_a$, where $n_a$ is the density of nuclear enzymes, $f_a$ is their effective dipolar force transmitted along the chromatin chain, and $\ell_a$ is the characteristic separation length of the dipole. Single-molecule assays provide measurements of stall forces for nuclear enzymes such as helicases, RNA polymerases (Pol\,II), and DNA polymerases~\cite{Bustamante2011, Wuite2000, Galburt2007}, but the effective force $f_a$ relevant here depends on how much of this load projects along the chromatin chain direction and may be reduced by geometric factors, local compliance, and load sharing. As a conservative working estimate based on Pol II, we take $f_a= 1\!-\!10~\mathrm{pN}$ and $\ell_a= 3\!-\!10~\mathrm{nm}$~\cite{Galburt2007, Abbondanzieri2005}. Experimentally, the density of active Pol\,II foci has been measured to be $\sim1~\mu\mathrm{m}^{-2}$~\cite{Zidovska2013}; by using a confocal depth of $\sim0.5~\mu\mathrm{m}$ and assuming $1$--$10$ Pol\,II per focus yields $n_a=2\!-\!20~\mu\mathrm{m}^{-3}$. With $\alpha_0=(n_a/\rho_E)f_a\ell_a$ and $\alpha=\alpha_0 R/\sqrt{\eta_s d_c B}$, we obtain $\alpha= 10^{2}\!-\!10^{5}$ (that is, $\alpha_0 = 10^{-4}$--$10^{-1}\,\mathrm{pN}\cdot\mathrm{nm}$), and the values used in our simulations lie toward the lower end of this range. Nevertheless, measurements for other active nuclear enzymes, such as DNA polymerases, report larger forces in the tens of pN~\cite{Wuite2000}, so the estimates of $\alpha$ based on Pol\,II should be regarded as a lower bound.}

     {Table~II lists all dimensional parameters in our model with their associated physiological ranges, some obtained from experiments, while others estimated indirectly from microscopic considerations.}

	\section{Linear stability analysis}	
	
	The homogeneous and stationary state of the governing equations in Table I is given by $\bar{\rho}_E$  and $\bar{\rho}_H$, together with vanishing velocities: $\bar{\boldsymbol{u}}_S= \boldsymbol{0}$, $\bar{\boldsymbol{u}}_E = \boldsymbol{0}$, and $\bar{\boldsymbol{u}}_H = \boldsymbol{0}$. We perturb the densities about their homogeneous state:
	\begin{equation}
		\label{eqn:density-perturb}
		\rho_{E} = 	\bar{\rho}_{E} + \varepsilon\,\delta\rho_E,\quad\text{and}\quad\rho_{E} = 	\bar{\rho}_{H} + \varepsilon\,\delta\rho_H,
	\end{equation} where $\delta\rho_E$ and $\delta\rho_H$ are the linear perturbations in the euchromatin and heterochromatin densities, respectively, while $\varepsilon$ is a perturbation parameter, which is immaterial purely keeping track of the perturbation order. Similarly, we perturb the velocities about their stationary state
	\begin{equation}
		\label{eqn:velocity-perturb}
		\boldsymbol{u}_{S} = \varepsilon\,\delta\boldsymbol{u}_S,\;\;\,
		\boldsymbol{u}_{E} = \varepsilon\,\delta\boldsymbol{u}_E,\;\;\,\text{and}\;\;\,\boldsymbol{u}_{H} = \varepsilon\,\delta\boldsymbol{u}_H.
	\end{equation}
	
	\subsection{Linearized chromatin density equations}
	
	To first order in $\varepsilon$, the continuity equations for the chromatin components can be written as follows:
	\begin{align}
		\label{eqn:lin-den-E}
		\frac{\partial}{\partial t}(\delta\rho_E) + \bar{\rho}_E\, \boldsymbol{\nabla}\cdot\delta\boldsymbol{u}_E &= \Delta(\delta\rho_E),\\[6pt]
		\label{eqn:lin-den-H}
		\frac{\partial}{\partial t}(\delta\rho_H) + \bar{\rho}_H\, \boldsymbol{\nabla}\cdot\delta\boldsymbol{u}_H &= \Delta(\delta\rho_H),
	\end{align} while the solvent incompressibility condition becomes
	\begin{equation}
		\label{eqn:lin-incomp-cond}
		\boldsymbol{\nabla}\cdot\delta\boldsymbol{u}_S = 0.
	\end{equation}
	
	Similarly, by expanding to first order in $\varepsilon$ the momentum equations of euchromatin and heterochromatin components, we find
	\begin{align}
		\label{eqn:lin-mom-uE}
		&\eta{\bar{\rho}_E}\big[\Delta\delta\boldsymbol{u}_E + \boldsymbol{\nabla}(\boldsymbol{\nabla}\!\cdot\delta\boldsymbol{u}_E)\big]\! + \Gamma\bar{\rho}_E(\delta\boldsymbol{u}_S-\delta\boldsymbol{u}_E) \notag\\[6pt]
		&\quad-\boldsymbol{\nabla}\Big[\frac{\bar{\rho}_E}{2}\delta\rho_H + \Big(\bar{\rho}_E + \frac{\bar{\rho}_H}{2}\Big)\delta{\rho}_E\Big]\! =  0,\\[14pt]
		\label{eqn:lin-mom-uH}
		&\eta{\bar{\rho}_H}\big[\Delta\delta\boldsymbol{u}_H + \boldsymbol{\nabla}(\boldsymbol{\nabla}\!\cdot\delta\boldsymbol{u}_H)\big]\! + \Gamma\bar{\rho}_H(\delta\boldsymbol{u}_S-\delta\boldsymbol{u}_H) \; \notag\\[6pt]
		&\quad-\boldsymbol{\nabla}\Big[\frac{\bar{\rho}_H}{2}\delta\rho_E
		+ \Big(\bar{\rho}_H-\bar{\rho}_C + \frac{\bar{\rho}_E}{2}\Big)\delta{\rho}_H\Big]\! =  0,
	\end{align}  respectively. By applying the divergence operator onto Eq.~(\ref{eqn:lin-mom-uE}) and (\ref{eqn:lin-mom-uH}), we obtain the following equations:   
	\begin{align}
		\label{eqn:lin-mom-E}
		2\eta{\bar{\rho}_E}\Delta( \boldsymbol{\nabla}\cdot\delta\boldsymbol{u}_E) 
		&= \Gamma \bar{\rho}_E(\boldsymbol{\nabla}\cdot\delta\boldsymbol{u}_E) + \frac{\bar{\rho}_E}{2}\Delta(\delta\rho_H) \notag\\[5pt] 
		& \quad+\left[\bar{\rho}_E+\frac{\bar{\rho}_H}{2}\right]\!\Delta(\delta\rho_E) ,\\[12pt]
		\label{eqn:lin-mom-H}
		2\eta{\bar{\rho}_H}\Delta( \boldsymbol{\nabla}\cdot\delta\boldsymbol{u}_H) 
		&= \Gamma \bar{\rho}_H(\boldsymbol{\nabla}\cdot\delta\boldsymbol{u}_H) + \frac{\bar{\rho}_H}{2}\Delta(\delta\rho_E)\notag\\[5pt] 
		& \quad+\left[\bar{\rho}_H - \bar{\rho}_C + \frac{\bar{\rho}_E}{2}\right]\Delta(\delta\rho_H).
	\end{align} Therefore, the divergence terms $\boldsymbol{\nabla}\cdot\delta\boldsymbol{u}_E$ and $\boldsymbol{\nabla}\cdot\delta\boldsymbol{u}_H$ can be expressed solely in terms of density fields by means of Eq.~(\ref{eqn:lin-den-E}) and (\ref{eqn:lin-den-H}), respectively, which by substitution into Eq.~(\ref{eqn:lin-mom-E}) and (\ref{eqn:lin-mom-H}) leads to
	\begin{align}
		\label{eqn:lin-rel-E}
		\!\!\frac{\partial}{\partial t}{\left(\Gamma\!- {2\eta}\Delta\right)} \delta\rho_E &= \left[\Gamma+\bar{\rho}_E+\frac{\bar{\rho}_H}{2}\right]\!\Delta(\delta\rho_E) \notag \\[5pt]
		&\;+\frac{\bar{\rho}_E}{2}\Delta(\delta\rho_H)-{2\eta}\Delta^2(\delta\rho_E),\\[10pt]
		\label{eqn:lin-rel-H}
		\frac{\partial}{\partial t}{\left(\Gamma\! - {2\eta}\Delta\right)} \delta\rho_H &= \left[\Gamma+\bar{\rho}_H - \bar{\rho}_C+\frac{\bar{\rho}_E}{2}\right]\!\Delta(\delta\rho_H) \notag \\[5pt]
		&\;+\frac{\bar{\rho}_H}{2}\Delta(\delta\rho_E)-{2\eta}\Delta^2(\delta\rho_H),
	\end{align} where $\Delta^2$ is the biharmonic operator.
	
	By expressing the linearized densities $\delta\rho_E$ and $\delta\rho_H$ in Fourier series, namely
	\begin{align}
		\delta\rho_E(\boldsymbol{r},t) = \sum_{\boldsymbol{k}}\,e^{i\boldsymbol{k}\cdot\boldsymbol{r}}\,\delta\hat{\rho}_E^{\,\boldsymbol{k}}(t),\\[8pt]
		\delta\rho_H(\boldsymbol{r},t) = \sum_{\boldsymbol{k}}\,e^{i\boldsymbol{k}\cdot\boldsymbol{r}}\,\delta\hat{\rho}_H^{\,\boldsymbol{k}}(t),
	\end{align} where $\boldsymbol{k}$ is the Fourier wave-vector and $\delta\hat{\rho}_{E,H}^{\,\boldsymbol{k}}$ are the corresponding Fourier amplitudes of $\delta\rho_{E,H}$, we can rewrite   Eq.~(\ref{eqn:lin-rel-E}) and (\ref{eqn:lin-rel-H}) solely in terms of $\delta\hat{\rho}_{E}^{\,\boldsymbol{k}}$ and $\delta\hat{\rho}_{H}^{\,\boldsymbol{k}}$. This can written in the following matrix form:
	\begin{equation}
		\frac{\partial}{\partial t}\!\begin{bmatrix}
			\delta\hat{\rho}_{E}^{\,\boldsymbol{k}} \\[3pt]
			\delta\hat{\rho}_{H}^{\,\boldsymbol{k}} 
		\end{bmatrix} =	 \,\boldsymbol{\mathcal{S}}_{\hspace{-0.5pt}{k}}\!\begin{bmatrix}\delta\hat{\rho}_{E}^{\,\boldsymbol{k}} \\[3pt]
			\delta\hat{\rho}_{H}^{\,\boldsymbol{k}} 
		\end{bmatrix}\!,
	\end{equation} where the matrix $\boldsymbol{\mathcal{S}}_{\hspace{-0.5pt}{k}}$ is shown in Eq.~(\ref{eqn:stability-matrix}), and $k$ is the magnitude of $\boldsymbol{k}$. The eigenvalues of $\boldsymbol{\mathcal{S}}_{\hspace{-0.5pt}{k}}$ are the growth rates of the Fourier amplitudes of density perturbations. These eigenvalues are found to be
	\begin{equation}
		\label{eqn:density-eigenvalues}
		\lambda_\pm(k) = -\frac{\,k^2}{2}\!\left[a_k+b_k\pm\sqrt{c_k + (a_k-b_k)^2}\,\right]
	\end{equation} where $a_k$ and $b_k$ are diagonal components of the matrix in Eq.~(\ref{eqn:stability-matrix}), that is, $a_k = k^{-2}[\boldsymbol{\mathcal{S}}_{\hspace{-0.5pt}{k}}]_{1,1}$ and $b_k = k^{-2}[\boldsymbol{\mathcal{S}}_{\hspace{-0.5pt}{k}}]_{2,2}$, whilst $c_k = 4k^{-4}[\boldsymbol{\mathcal{S}}_{\hspace{-0.5pt}{k}}]_{1,2}[\boldsymbol{\mathcal{S}}_{\hspace{-0.5pt}{k}}]_{2,1}$. The solution $\lambda_+$ is always negative, while the other could become positive, displaying a band of linearly unstable modes. We denote the latter by $\lambda_C$. The instability condition is determined by the sign of the growth rate $\lambda_C(k)$ near $k=0$; namely,
	\begin{equation*}
		\lambda_C = \frac{\frac{2\bar{\rho}_C-4\Gamma}{\bar{\rho}_E+\bar{\rho}_H}\!- 3 +\! \sqrt{1\!+\!\frac{4\bar{\rho}_C^2+4\bar{\rho}_C(\bar{\rho}_E-\bar{\rho}_H)}{(\bar{\rho}_E+\bar{\rho}_H)^2}}}{\frac{4\Gamma}{\bar{\rho}_E+\bar{\rho}_H}}\,k^2 +\mathcal{O}[k^3].
	\end{equation*} We find that the system remains stable to linear density perturbations whenever
	\begin{equation}
		\bar{\rho}_C\le\bar{\rho}^{\,\star}_C = \frac{(\Gamma + \bar{\rho}_E+\bar{\rho}_H)(2\Gamma + \bar{\rho}_E+\bar{\rho}_H)}{2(\Gamma+\rho_E)+\bar{\rho}_H},
	\end{equation} with the equality corresponding to the onset of instability. The system is linearly unstable if $\bar{\rho}_C>\bar{\rho}^{\,\star}_C$, which is shown in Fig.~\ref{fig:figures/figure_2}(b). Note that this criterion is independent of the viscosity ratio $\eta$. However the fastest growing mode $k_{0}$ satisfies $\lambda_C'(k_{0})=0$ and depends on viscosity as
	\begin{equation}
		k_{0}\propto{\eta^{-1/2}}.
	\end{equation} This scaling is readily obtain by noting that $\lambda_C'(k)=0$ can be rewritten as a polynomial in terms of $k\sqrt{\eta}$. 
	
	Also, note that this instability is completely independent of the activity parameter $\alpha$, as the extensile active stresses act only on the solvent (nucleoplasm), which is treated here as an incompressible fluid.
	
	\subsection{Linearized solvent equations}
	
	Lastly, we also expand to linear order in $\varepsilon$ the momentum equation of the solvent, using Eq.~(\ref{eqn:density-perturb}) and (\ref{eqn:velocity-perturb}), and a perturbation in the nematic order parameter,
	\begin{equation}
		\label{eqn:Q-perturb}
		\boldsymbol{Q} = 	\bar{\boldsymbol{Q}} + \varepsilon\,\delta\boldsymbol{Q}(\boldsymbol{r},t),
	\end{equation}  about a homogeneous state $\bar{\boldsymbol{Q}}$, where $\delta\boldsymbol{Q}$ are the corresponding linear perturbations. This leads to the following expansion to first order in $\varepsilon$:
	\begin{align}
		\label{eqn:lin-mom-S}
		\Delta(\delta\boldsymbol{u}_S)-\!\boldsymbol{\nabla} \delta p = &\,\Gamma\bar{\rho}_H(\delta\boldsymbol{u}_S-\delta\boldsymbol{u}_H)+\Gamma\bar{\rho}_E(\delta\boldsymbol{u}_S-\delta\boldsymbol{u}_E)\notag\\[5pt]
		&+\boldsymbol{\nabla}\!\cdot\!(\alpha\bar{\rho}_E\delta\boldsymbol{Q}) + \alpha\,\bar{\boldsymbol{Q}}\cdot\boldsymbol{\nabla}\delta\rho_E,
	\end{align} where the solvent pressure $p = \bar{p} + \varepsilon\,\delta p(\boldsymbol{r},t)$. To zeroth order in $\varepsilon$, any constant and symmetric tensor $\bar{\boldsymbol{Q}}$ satisfies Eq.~(\ref{eqn:Q-dyanmics}). To first order in $\varepsilon$, the rescaled nematic order equation (see Table I) becomes
	\begin{align}
		\label{eqn:lin-Q-dyn}
		\frac{\partial}{\partial t}(\delta\boldsymbol{Q})&=\bar{\boldsymbol{Q}}\cdot(\boldsymbol{\nabla}\delta\boldsymbol{u}_S)+(\boldsymbol{\nabla}\delta\boldsymbol{u}_S)^{\mathsf{T}}\cdot\bar{\boldsymbol{Q}} \notag\\[2pt] 
		&\quad + \Delta(\delta\boldsymbol{Q})-2\hspace{1pt}\bar{\boldsymbol{Q}}\,\bar{\boldsymbol{Q}}:\boldsymbol{\nabla}\delta\boldsymbol{u}_S.
	\end{align} 
	
	By Fourier transforming the linearized solvent momentum equation in (\ref{eqn:lin-mom-S}), we find 
	\begin{align}
		\label{eqn:lin-mom-S-Fourier}
		\!\!i\boldsymbol{k}\delta \hat{p}_{\boldsymbol{k}} +  k^2\delta\hat{\boldsymbol{u}}^{\boldsymbol{k}}_S  = &\;\Gamma\bar{\rho}_H(\delta\hat{\boldsymbol{u}}^{\boldsymbol{k}}_H-\delta\hat{\boldsymbol{u}}^{\boldsymbol{k}}_S)+\Gamma\bar{\rho}_E(\delta\hat{\boldsymbol{u}}^{\boldsymbol{k}}_E-\delta\hat{\boldsymbol{u}}^{\boldsymbol{k}}_S)\notag\\[5pt]
		&\!-i\boldsymbol{k}\!\cdot\!(\alpha\bar{\rho}_E\delta\hat{\boldsymbol{Q}}_{\boldsymbol{k}}) - \alpha\,\bar{\boldsymbol{Q}}\cdot (i{\boldsymbol{k}}\delta\hat{\rho}^{\boldsymbol{k}}_E),
	\end{align} while the incompressibility condition in Eq.~(\ref{eqn:lin-incomp-cond}) becomes
	\begin{equation}
		\label{eqn:lin-incomp-cond-k}
		\boldsymbol{k}\cdot\delta\hat{\boldsymbol{u}}^{\boldsymbol{k}}_S = 0.
	\end{equation}
	
	Similarly, linearized dynamics of the nematic order parameter in Eq.~(\ref{eqn:lin-Q-dyn}) can be written in Fourier space as:
	\begin{align}
		\label{eqn:lin-Q-dyn-Fourier}
		\frac{\partial}{\partial t}(\delta\hat{\boldsymbol{Q}}_{\boldsymbol{k}})&=\bar{\boldsymbol{Q}}\cdot(i\boldsymbol{k}\,\delta\hat{\boldsymbol{u}}^{\boldsymbol{k}}_S)+(i\boldsymbol{k}\,\delta\hat{\boldsymbol{u}}^{\boldsymbol{k}}_S)^{\mathsf{T}}\cdot\bar{\boldsymbol{Q}} \notag\\[2pt] 
		&\quad - k^2(\delta\hat{\boldsymbol{Q}}_{\boldsymbol{k}})-2\hspace{1pt}\bar{\boldsymbol{Q}}\,\bar{\boldsymbol{Q}}:(i\boldsymbol{k}\,\delta\hat{\boldsymbol{u}}^{\boldsymbol{k}}_S),
	\end{align} where $\delta \hat{p}_{\boldsymbol{k}}(t)$ and $\delta \hat{\boldsymbol{Q}}_{\boldsymbol{k}}(t)$ are Fourier amplitudes of the linearized pressure $\delta p$ and linearized nematic order parameter $\delta\boldsymbol{Q}$, respectively. Similarly,  $\delta\hat{\boldsymbol{u}}^{\boldsymbol{k}}_S(t)$, $\delta\hat{\boldsymbol{u}}^{\boldsymbol{k}}_E(t)$, and $\delta\hat{\boldsymbol{u}}^{\boldsymbol{k}}_H(t)$ are Fourier amplitudes of the linearized velocity fields of solvent, euchromatin and heterochromatin, respectively. The latter two velocity amplitudes, $\delta\hat{\boldsymbol{u}}^{\boldsymbol{k}}_E$ and $\delta\hat{\boldsymbol{u}}^{\boldsymbol{k}}_H$, are determined from the Fourier transform of momentum equations in (\ref{eqn:lin-mom-uE}) and (\ref{eqn:lin-mom-uH}), namely 
	\begin{align}
		\label{eqn:lin-mom-uE-k}
		\!\eta{\bar{\rho}_E}\big[&k^2\delta\hat{\boldsymbol{u}}^{\boldsymbol{k}}_E + \boldsymbol{k}(\boldsymbol{k}\cdot\delta\hat{\boldsymbol{u}}^{\boldsymbol{k}}_E)\big]\! + \Gamma\bar{\rho}_E(\delta\hat{\boldsymbol{u}}^{\boldsymbol{k}}_E-\delta\hat{\boldsymbol{u}}^{\boldsymbol{k}}_S)\;
		\notag\\[6pt]
		&+i\boldsymbol{k}\Big[\frac{\bar{\rho}_E}{2}\delta\hat{\rho}^{\,\boldsymbol{k}}_H + \!\Big(\bar{\rho}_E + \frac{\bar{\rho}_H}{2}\Big)\delta{\hat{\rho}}^{\,\boldsymbol{k}}_E\Big]\! =  0,\\[12pt]
		\label{eqn:lin-mom-uH-k}
		\!\eta{\bar{\rho}_H}\big[&k^2\delta\hat{\boldsymbol{u}}^{\boldsymbol{k}}_H + \boldsymbol{k}(\boldsymbol{k}\cdot\delta\hat{\boldsymbol{u}}^{\boldsymbol{k}}_H)\big]\! + \Gamma\bar{\rho}_H(\delta\hat{\boldsymbol{u}}^{\boldsymbol{k}}_H-\delta\hat{\boldsymbol{u}}^{\boldsymbol{k}}_S)\;
		\notag\\[6pt]
		&+i\boldsymbol{k}\Big[\frac{\bar{\rho}_H}{2}\delta\hat{\rho}^{\,\boldsymbol{k}}_E + \!\Big(\bar{\rho}_H -\bar{\rho}_{C} + \frac{\bar{\rho}_E}{2}\Big)\delta{\hat{\rho}}^{\,\boldsymbol{k}}_H\Big]\! =  0.
	\end{align}

	From equations (\ref{eqn:lin-mom-S-Fourier}), (\ref{eqn:lin-incomp-cond-k}), (\ref{eqn:lin-mom-uE-k}), and (\ref{eqn:lin-mom-uH-k}), we can analytically determine the pressure $\delta \hat{p}_{\boldsymbol{k}}$ and the solvent velocity $\delta\hat{\boldsymbol{u}}^{\boldsymbol{k}}_S$, as well as the euchromatin velocity $\delta\hat{\boldsymbol{u}}^{\boldsymbol{k}}_E$ and the heterochromatin velocity $\delta\hat{\boldsymbol{u}}^{\boldsymbol{k}}_H$, purely in terms of the  densities $\delta\hat{\rho}_E^{\,\boldsymbol{k}}$ and $\delta\hat{\rho}_H^{\,\boldsymbol{k}}$. Thus, the linearized dynamics of the order parameter in Eq.~(\ref{eqn:lin-Q-dyn-Fourier}), together with continuity equations in Fourier space, that is,
	\begin{align}
		\label{eqn:lin-den-E-k}
		\frac{\partial}{\partial t}(\delta\hat{\rho}^{\,\boldsymbol{k}}_E) + \bar{\rho}_E (\boldsymbol{k}\cdot\delta\hat{\boldsymbol{u}}^{\boldsymbol{k}}_E) &= -k^2\delta\hat{\rho}^{\,\boldsymbol{k}}_E,\\[6pt]
		\label{eqn:lin-den-H-k}
		\frac{\partial}{\partial t}(\delta\hat{\rho}^{\,\boldsymbol{k}}_H) + \bar{\rho}_H (\boldsymbol{k}\cdot\delta\hat{\boldsymbol{u}}^{\boldsymbol{k}}_H) &= -k^2\delta\hat{\rho}^{\,\boldsymbol{k}}_H,
	\end{align} forms a closed set of linear differential equations in $\delta\hat{\rho}_E^{\,\boldsymbol{k}}$, $\delta\hat{\rho}_H^{\,\boldsymbol{k}}$ and the independent components of $\delta\hat{\boldsymbol{Q}}_{\boldsymbol{k}}$ (specifically, 3 in two dimensions, and 6 in three dimensions).
	
	\subsection{Nematic order parameter}

	The base state of $\boldsymbol{Q}$ is a symmetric, constant tensor, but its components are not entirely unconstrained, as 
	\begin{equation}
		\label{eqn:Q-def}
		{\boldsymbol{Q}} = \frac{\int\boldsymbol{p}\boldsymbol{p}\,{\Psi}(\boldsymbol{p})\;\mathrm{d}^d\boldsymbol{p}}{\int{\Psi}(\boldsymbol{p})\;\mathrm{d}^d\boldsymbol{p}},
	\end{equation} where $\boldsymbol{p}$ is a unit vector in dimension $d$, ${\Psi}$ is an orientational distribution, and the integration is performed  over the $d$-dimensional unit sphere~\cite{DoiBook1986, Saintillan2008, Saintillan2013, Gao2017[Rev]}. The definition in Eq.~(\ref{eqn:Q-def}) implies that the trace of $\boldsymbol{Q}$ is unity, 
	$\mathrm{tr}(\boldsymbol{Q}) = 1$. We denote the distribution at zeroth order in the perturbation as $\bar{\Psi}$, which can described by either a constant,
	\begin{equation}
		\bar{\Psi}(\boldsymbol{p}) = \Psi_0\quad\text{(isotropic state)},
	\end{equation} or a fixed orientation $\bar{\boldsymbol{n}}$, that is,
	\begin{equation}
		\bar{\Psi}(\boldsymbol{p}) = \delta(\boldsymbol{p} - \bar{\boldsymbol{n}}),\quad\text{(aligned state)}
	\end{equation}  with $\delta$ as the Dirac--delta function. Specifically, we find 
	\begin{equation}
		\bar{\boldsymbol{Q}}_I\! = {d}^{-1}\boldsymbol{I}~\,\text{(isotropic)},\;\text{and}\;\, \bar{\boldsymbol{Q}}_A\! = \bar{\boldsymbol{n}}\bar{\boldsymbol{n}}~\,\text{(aligned)},
	\end{equation} where $\boldsymbol{I}$ is the identity tensor. In two dimensions,
	\begin{equation}
		\label{eqn:base-state-2D}
		\bar{\boldsymbol{Q}}_I = {\footnotesize\begin{bmatrix}
				\frac{1}{2} & 0\\
				0 & \frac{1}{2}\end{bmatrix}}, \quad\text{and}\quad\bar{\boldsymbol{Q}}_A = {\footnotesize\begin{bmatrix}
				0 & 0\\
				0 & 1
		\end{bmatrix}},
	\end{equation} where we choose $\bar{\boldsymbol{n}}$ to be aligned with the $y$-axis, without any loss of generality. Similarly, in three dimensions,
	\begin{equation}
		\label{eqn:base-state-3D}
		\bar{\boldsymbol{Q}}_I = {\footnotesize\begin{bmatrix}
				\frac{1}{3} & 0 & 0\\
				0 & \frac{1}{3} & 0\\
				0 & 0 & \frac{1}{3}\\
		\end{bmatrix}},\quad\text{and}\quad\bar{\boldsymbol{Q}}_A = {\footnotesize\begin{bmatrix}
				0 & 0 & 0\\
				0 & 0 & 0\\
				0 & 0 & 1\\
		\end{bmatrix}},
	\end{equation} where in the latter we choose $\bar{\boldsymbol{n}}$ to be along the $z$-axis.
    
	\begin{figure*}[t]\includegraphics[width=\textwidth]{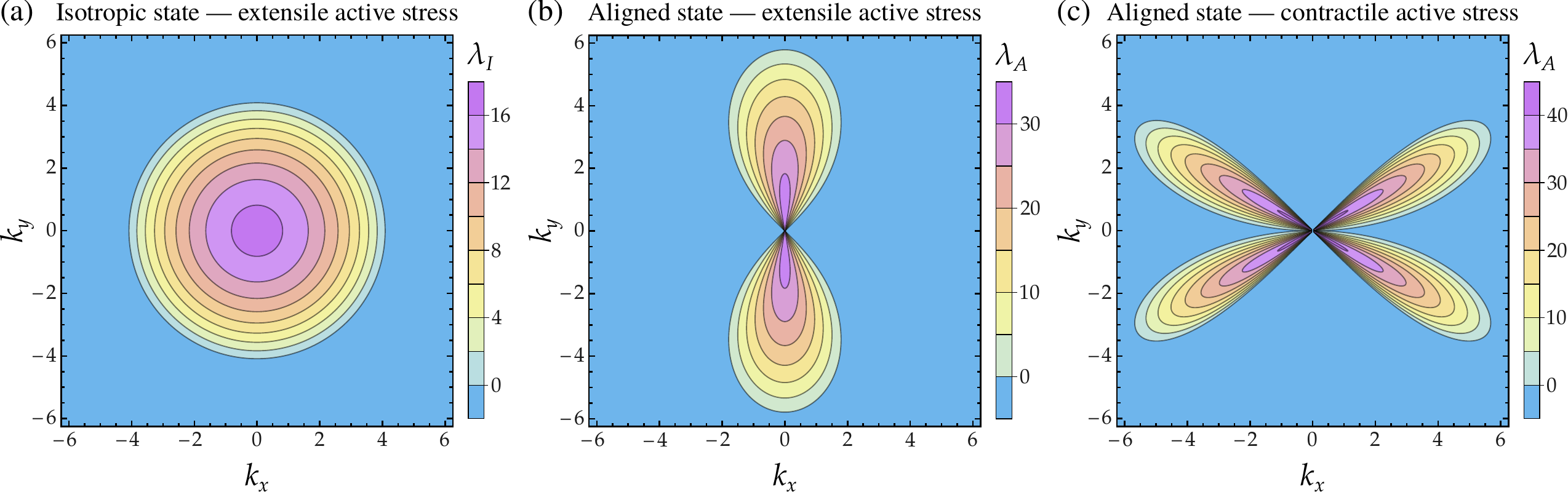}
		\caption{\label{fig:figures/figure_7} Diagrams showing instability regions in two dimensions as a function of components of the Fourier  vector $\boldsymbol{k}=[k_x,\,k_y]$, where the color shows only the positive values of the growth rates associated with linear perturbations in the nematic order parameter $\boldsymbol{Q}$ about: (a) an isotropic state for extensile active stresses with {$\alpha=5$}; (b) an aligned state along the $y$-direction, for extensile active stresses with {$\alpha=5$}; (c) an aligned state along the $y$-direction, for contractile active stresses with {$\alpha=-50$}. Here, we  choose {$\eta=0.01$ and $\Gamma=25$}, while the homogeneous densities {$\bar{\rho}_E = 10$ and  $\bar{\rho}_H=40$}. 
	}\end{figure*} 
    
	In two dimensions, the first-order perturbation in nematic order parameter $\delta\boldsymbol{Q}$ and its corresponding Fourier transform can be written as
	\begin{equation*}
		\delta{\boldsymbol{Q}} = {\footnotesize\begin{bmatrix}
				\delta q_1 & \delta q_3\\
				\delta q_3 & -\delta q_1
		\end{bmatrix}},\quad
		\delta{\hat{\boldsymbol{Q}}_{\boldsymbol{k}}} = {\footnotesize\begin{bmatrix}
				\delta\hat{q}^{\boldsymbol{k}}_1 & \delta\hat{q}^{\boldsymbol{k}}_3\\
				\delta\hat{q}^{\boldsymbol{k}}_3 & -\delta\hat{q}^{\boldsymbol{k}}_1
		\end{bmatrix}},
	\end{equation*} where $\delta q_j(\boldsymbol{r},t)$ and $\delta \hat{q}^{\hspace{1pt}\boldsymbol{k}}_j(t)$ are their respective independent components. Note that the trace of $\delta{\boldsymbol{Q}}$ must be zero such that we maintain $\mathrm{tr}(\boldsymbol{Q})=1$. Thus, the linear system associated with perturbations about the two-dimensional isotropic state $\bar{\boldsymbol{Q}}_I$ in Eq.~(\ref{eqn:base-state-2D}) is given by
	\begin{equation}
		\frac{\partial}{\partial t}\!{\footnotesize\begin{bmatrix}
				\delta\hat{q}^{\boldsymbol{k}}_1\\[1pt]
				\delta\hat{q}^{\boldsymbol{k}}_2\\[1pt]
				\delta \hat{\rho}^{\,\boldsymbol{k}}_E\\[1pt]
				\,\delta \hat{\rho}^{\,\boldsymbol{k}}_H\,
		\end{bmatrix}}\!= {\begin{bmatrix}
				\boldsymbol{\mathcal{H}}_I^{\boldsymbol{k}} & \boldsymbol{O}_{2\times2}\\[1pt]
				\,\boldsymbol{O}_{2\times2} & \boldsymbol{\mathcal{S}}_k
		\end{bmatrix}}{\footnotesize\begin{bmatrix}
				\delta\hat{q}^{\boldsymbol{k}}_1\\[1pt]
				\delta\hat{q}^{\boldsymbol{k}}_2\\[1pt]
				\delta \hat{\rho}^{\,\boldsymbol{k}}_E\\[1pt]
				\,\delta \hat{\rho}^{\,\boldsymbol{k}}_H\,
		\end{bmatrix}},
	\end{equation} where $\boldsymbol{\mathcal{S}}_k$ is the matrix from Eq.~(\ref{eqn:stability-matrix}), $\boldsymbol{O}_{a\times b}$ is a zero matrix of $a$ rows and $b$ columns, and $\boldsymbol{\mathcal{H}}_I^{\boldsymbol{k}}$ is found to be
	\begin{equation}
		\boldsymbol{\mathcal{H}}_I^{\boldsymbol{k}} = -k^2 \boldsymbol{I}_{2\times2} + \frac{1}{2}\Lambda(k) \,\boldsymbol{\mathcal{R}}_I^{\boldsymbol{k}},
	\end{equation}  where $\boldsymbol{I}_{2\times2}$ is the identity matrix, the function  
	\begin{equation}
		\label{eqn:Lambda-def}
		{\Lambda(k) = \frac{\alpha\bar{\rho}_E}{\displaystyle 1+\frac{\eta\,\Gamma\left(\bar{\rho}_E+\bar{\rho}_H\right)}{\Gamma+\eta\,k^2}}},
	\end{equation} and the matrix 
	\begin{equation}
		\boldsymbol{\mathcal{R}}_I^{\boldsymbol{k}} =\!{\footnotesize\begin{bmatrix}
				\sin ^2(2 \varphi ) & -\frac{1}{2} \sin (4 \varphi ) \\
				-\frac{1}{2} \sin (4 \varphi ) & \cos ^2(2 \varphi ) \\
		\end{bmatrix}}\!,
	\end{equation} with polar angle $\varphi$ defining the components $k_x = k\cos\varphi$ and $k_y = k\sin\varphi$ of the wavevector $\boldsymbol{k} = [k_x,\,k_y]$. One of the eigenvalues of $\boldsymbol{\mathcal{H}}_I^{\boldsymbol{k}}$ is given by $-k^2$, while the other is
	\begin{equation}
		\lambda_I(k) = -k^2+\Lambda(k)/2,
	\end{equation} which shows that the isotropic state is always stable for the contractile case ($\alpha<0$), whereas for the extensile case ($\alpha>0$) it is always unstable; see Fig.~\ref{fig:figures/figure_7}(a).
	
	Similarly, the linear system associated with perturbations about an aligned state $\bar{\boldsymbol{Q}}_A$  is given by
	\begin{equation}
		\label{eqn:stability-matrix-A}
		\frac{\partial}{\partial t}\!{\footnotesize\begin{bmatrix}
				\delta\hat{q}^{\boldsymbol{k}}_1\\[1pt]
				\delta\hat{q}^{\boldsymbol{k}}_2\\[1pt]
				\delta \hat{\rho}^{\,\boldsymbol{k}}_E\\[1pt]
				\,\delta \hat{\rho}^{\,\boldsymbol{k}}_H\,
		\end{bmatrix}}\!= {\begin{bmatrix}\boldsymbol{\mathcal{H}}_A^{\boldsymbol{k}} & \boldsymbol{\mathcal{C}}_{\boldsymbol{k}}\\[1pt]
				\,\boldsymbol{O}_{2\times2} & \boldsymbol{\mathcal{S}}_k\end{bmatrix}}{\footnotesize\begin{bmatrix}
				\delta\hat{q}^{\boldsymbol{k}}_1\\[1pt]
				\delta\hat{q}^{\boldsymbol{k}}_2\\[1pt]
				\delta \hat{\rho}^{\,\boldsymbol{k}}_E\\[1pt]
				\,\delta \hat{\rho}^{\,\boldsymbol{k}}_H\,
		\end{bmatrix}},
	\end{equation}
	\begin{equation}
		\boldsymbol{\mathcal{H}}_A^{\boldsymbol{k}} = -k^2 \boldsymbol{I}_{2\times2}+\Lambda(k)\boldsymbol{\mathcal{R}}_A^{\boldsymbol{k}},
	\end{equation} with matrices
	\begin{align}
		\boldsymbol{\mathcal{R}}_A^{\boldsymbol{k}} &=\sin ^2(\varphi ){\footnotesize\begin{bmatrix}
				0 & 0 \\
				\sin (2\varphi ) & -\cos (2 \varphi ) \\
		\end{bmatrix}}\!,\\[8pt]
		\boldsymbol{\mathcal{C}}_{\boldsymbol{k}} &=-\frac{\Lambda(k)\cos (\varphi)\sin^3(\varphi )}{\bar{\rho}_E} {\footnotesize\begin{bmatrix}
				0 & 0 \\
				1 & 0 \\
		\end{bmatrix}}.
	\end{align} 
	
	In addition to the eigenvalues associated with density perturbations, as derived in Eq.~(\ref{eqn:density-eigenvalues}), the remaining eigenvalues of the stability matrix in Eq.~(\ref{eqn:stability-matrix-A}) are given by a purely diffusive eigenvalue, $-k^2$, and 
	\begin{equation}
		\lambda_A(k) = -k^2-\Lambda(k)\sin^2(\varphi)\cos(2\varphi).
	\end{equation} 
	
	The region of instability is given by $\lambda_A>0$, and shown in Fig.~\ref{fig:figures/figure_7}(b) and (c) for the extensile ($\alpha>0$) and contractile ($\alpha<0$) stresses, respectively. It can be shown that the fastest growing mode in the the contractile case is found when the angle $\varphi = \frac{\pi}{6}$, $\frac{5\pi}{6}$, $\frac{7\pi}{6}$, and  $\frac{11\pi}{6}$, while $k$ satisfies the equation $\Lambda'(k)\!+16k = 0$. In the case of extensile stresses, the fastest growing mode is given by angle $\varphi = \frac{\pi}{2}$ and $\frac{3\pi}{2}$, with $k$ satisfying  $\Lambda'(k)-2k=0$.

	\begin{figure*}[t!]\includegraphics[width=\textwidth]{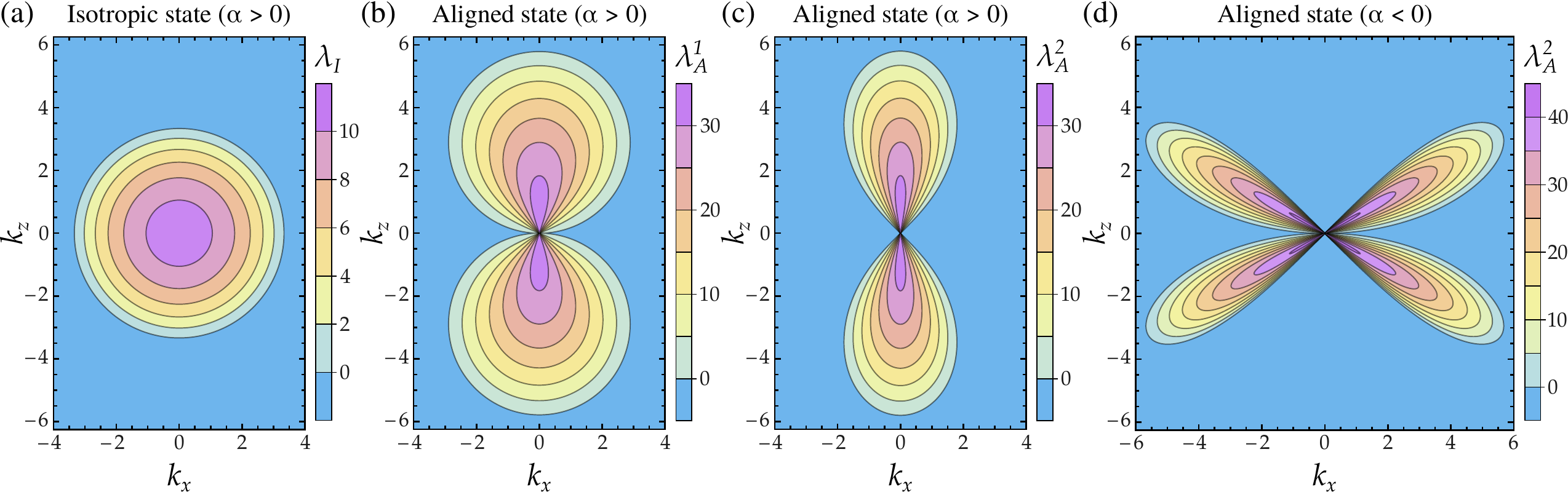}
		\caption{\label{fig:figures/figure_8} Diagrams of instability regions in three dimensions as a function of components of the Fourier vector $\boldsymbol{k}=[k_x,\,k_y,\,k_z]$, showing cross-sectional cuts at $k_y=0$. The regions are rotationally symmetric around the $k_z$--axis. Here, the color shows only the positive values of the growth rates associated with linear perturbations in the nematic order parameter $\boldsymbol{Q}$. (a) The unstable growth rate $\lambda_I$ about an isotropic state for extensile active stresses with {$\alpha=5$}. (b) and (c) shows unstable growth rates $\lambda_A^1$ and $\lambda_A^2$, respectively, about an aligned state along the $z$-direction, for extensile active stresses with {$\alpha=5$}. (d) The unstable growth rate $\lambda_A^2$ for perturbations about an aligned state along the $z$-direction, for contractile active stresses with {$\alpha=-50$} (note that $\lambda_A^1<0$ for $\alpha<0$). Here, we  choose {$\eta=0.01$ and $\Gamma=25$}, while the homogeneous chromatin densities {$\bar{\rho}_E = 10$ and  $\bar{\rho}_H=40$}. 
	}\end{figure*} 
	
	In three dimensions, the first-order perturbation in nematic order parameter $\delta\boldsymbol{Q}$ and its corresponding Fourier transform can be written as
	\begin{equation*}
		\delta{\boldsymbol{Q}} =\! {\footnotesize\begin{bmatrix}
				\delta q_1 & \delta q_3 & \delta q_4 \\
				\delta q_3 & \delta q_2 & \delta q_5 \\
				\delta q_4 & \delta q_5 & \!\!-\delta q_1\! - \delta q_2\!
		\end{bmatrix}}\!,\;
		\delta{\hat{\boldsymbol{Q}}_{\boldsymbol{k}}} = \!{\footnotesize\begin{bmatrix}
				\delta\hat{q}^{\boldsymbol{k}}_1 & \delta\hat{q}^{\boldsymbol{k}}_3 & \delta\hat{q}^{\boldsymbol{k}}_4\\
				\delta\hat{q}^{\boldsymbol{k}}_3 & \delta\hat{q}^{\boldsymbol{k}}_2 & \delta\hat{q}^{\boldsymbol{k}}_5 \\
				\delta\hat{q}^{\boldsymbol{k}}_4 & \delta\hat{q}^{\boldsymbol{k}}_5 & \!\!-\delta\hat{q}^{\boldsymbol{k}}_1-\delta\hat{q}^{\boldsymbol{k}}_2\! \\
		\end{bmatrix}}\!,
	\end{equation*} where $\delta q_j(\boldsymbol{r},t)$ and $\delta \hat{q}^{\hspace{1pt}\boldsymbol{k}}_j(t)$ are  independent components.  The linear system associated with perturbations about the three-dimensional isotropic state $\bar{\boldsymbol{Q}}_I$ in Eq.~(\ref{eqn:base-state-3D}) is 
	\begin{equation}
		\frac{\partial}{\partial t}\!{\footnotesize\begin{bmatrix}
				\delta\hat{q}^{\boldsymbol{k}}_1\\
				\vdots\\
				\delta\hat{q}^{\boldsymbol{k}}_5\\[1pt]
				\delta \hat{\rho}^{\,\boldsymbol{k}}_E\\[1pt]
				\,\delta \hat{\rho}^{\,\boldsymbol{k}}_H\,
		\end{bmatrix}}\!= {\begin{bmatrix}
				\boldsymbol{\mathcal{M}}_I^{\boldsymbol{k}} & \boldsymbol{O}_{5\times2}\\[1pt]
				\,\boldsymbol{O}_{2\times5} & \boldsymbol{\mathcal{S}}_k
		\end{bmatrix}}{\footnotesize\begin{bmatrix}
				\delta\hat{q}^{\boldsymbol{k}}_1\\
				\vdots\\
				\delta\hat{q}^{\boldsymbol{k}}_5\\[1pt]
				\delta \hat{\rho}^{\,\boldsymbol{k}}_E\\[1pt]
				\,\delta \hat{\rho}^{\,\boldsymbol{k}}_H\,
		\end{bmatrix}},
	\end{equation} with
	\begin{equation}
		\boldsymbol{\mathcal{M}}_I^{\boldsymbol{k}} = -k^2 \boldsymbol{I}_{5\times5} + \frac{1}{3}\Lambda(k) \,\boldsymbol{\mathcal{U}}_I^{\boldsymbol{k}},
	\end{equation}  where $\boldsymbol{I}_{5\times5}$ is the five-dimensional identity matrix, $\Lambda(k)$ is defined in Eq.~(\ref{eqn:Lambda-def}), and  matrix $\boldsymbol{\mathcal{U}}_I^{\boldsymbol{k}}$ is given solely in terms of the angles $\varphi\in[0,2\pi]$ and $\vartheta\in[0,\pi]$ by expressing the components of $\boldsymbol{k} = [k_x,\,k_y,\,k_z]$ in spherical coordinates: $k_x = k\sin(\vartheta)\cos(\varphi)$, $k_y = k\sin(\vartheta)\sin(\varphi)$, and $k_z = k\cos(\vartheta)$. The expression of $\boldsymbol{\mathcal{U}}_I^{\boldsymbol{k}}$ is cumbersome and will not be shown here. Three of the eigenvalues of $\boldsymbol{\mathcal{M}}_I^{\boldsymbol{k}}$ are $-k^2$ and the other two are both given by
	\begin{equation}
		\lambda_I(k) = -k^2+\Lambda(k)/3.
	\end{equation} Again, as in two dimensions, we find the isotropic state to be always unstable (stable) for extensile (contractile) active stresses, see Fig.~\ref{fig:figures/figure_8}(a). 
	
	The linear system associated with perturbations about the aligned state $\bar{\boldsymbol{Q}}_A$ in Eq.~(\ref{eqn:base-state-3D}) reads as follows:
	\begin{equation}
		\frac{\partial}{\partial t}\!{\footnotesize\begin{bmatrix}
				\delta\hat{q}^{\boldsymbol{k}}_1\\
				\vdots\\
				\delta\hat{q}^{\boldsymbol{k}}_5\\[1pt]
				\delta \hat{\rho}^{\,\boldsymbol{k}}_E\\[1pt]
				\,\delta \hat{\rho}^{\,\boldsymbol{k}}_H\,
		\end{bmatrix}}\!= {\begin{bmatrix}
				\boldsymbol{\mathcal{M}}_A^{\boldsymbol{k}} & \boldsymbol{\mathcal{B}}_{\boldsymbol{k}}\\[1pt]
				\,\boldsymbol{O}_{2\times5} & \boldsymbol{\mathcal{S}}_k
		\end{bmatrix}}{\footnotesize\begin{bmatrix}
				\delta\hat{q}^{\boldsymbol{k}}_1\\
				\vdots\\
				\delta\hat{q}^{\boldsymbol{k}}_5\\[1pt]
				\delta \hat{\rho}^{\,\boldsymbol{k}}_E\\[1pt]
				\,\delta \hat{\rho}^{\,\boldsymbol{k}}_H\,
		\end{bmatrix}},
	\end{equation} 
	\begin{equation}
		\boldsymbol{\mathcal{M}}_A^{\boldsymbol{k}} = -k^2 \boldsymbol{I}_{5\times5}+\Lambda(k)\,\boldsymbol{\mathcal{U}}_A^{\boldsymbol{k}},
	\end{equation} with matrix $\boldsymbol{\mathcal{U}}_A^{\boldsymbol{k}}$ defined solely in terms of the angles $\vartheta$ and $\varphi$ (not shown), while the matrix $\boldsymbol{\mathcal{B}}_{\boldsymbol{k}}$ is given by
	\begin{equation}
		\boldsymbol{\mathcal{B}}_{\boldsymbol{k}} =-\frac{\Lambda(k)\cos^3(\vartheta)\sin(\vartheta )}{\bar{\rho}_E} {\footnotesize\begin{bmatrix}
				0 & 0\;\, \\
				0 & 0\;\, \\
				0 & 0\;\, \\
				\,\cos\varphi & 0\;\, \\
				\,\sin\varphi & 0\;\, \\
		\end{bmatrix}}.
	\end{equation} 
	
	We find that three eigenvalues of the matrix $\boldsymbol{\mathcal{M}}_A^{\boldsymbol{k}}$ are $-k^2$, whereas the remaining two are given by
	\begin{align}
		\label{eqn:lambdaA_1}
		\lambda^{\mathsf{(1)}}_A(k) &= -k^2+\Lambda(k)\cos^2(\vartheta),\\[2pt]
		\label{eqn:lambdaA_2}
		\lambda^{\mathsf{(2)}}_A(k) &= -k^2+\Lambda(k)\cos^2(\vartheta)\cos(2\vartheta).
	\end{align} As a result, if $\alpha>0$ (extensile case), then both eigenvalues $\lambda^{\mathsf{(1)}}_A$ and $\lambda^{\mathsf{(2)}}_A$ are strictly positive within the same range of $k_z$ modes, see Fig.~\ref{fig:figures/figure_8}(b) and (c). However, if $\alpha<0$ (contractile case), then $\lambda^{\mathsf{(1)}}_A<0$ for any $\boldsymbol{k}$ vector, and only the eigenvalue $\lambda^2_A$ can become strictly positive, resulting in a linear instability, as shown in Fig.~\ref{fig:figures/figure_8}(d). 
	
	The fastest growing mode for $\alpha>0$ is found when the angle $\vartheta=0$ and $\pi$, and $k$ satisfies $\Lambda'(k)-2k=0$ (as found in two dimensions).  Likewise, the fastest growing mode for the contractile case ($\alpha<0$) is found at the angle $\vartheta=\frac{\pi}{3}$, $\frac{2\pi}{3}$, $\frac{4\pi}{3}$, and  $\frac{5\pi}{3}$, where $k$ satisfies $\Lambda'(k)+16k=0$ (also as in two dimensions). Note that the eigenvectors associated with $\lambda^{\mathsf{(1)}}_A$ and $\lambda^{\mathsf{(2)}}_A$ in Eqs.~(\ref{eqn:lambdaA_1}) and (\ref{eqn:lambdaA_2}) are found to be 
	\begin{align}
		\boldsymbol{e}^{\mathsf{(1)}}_A &= [0,0,0,-\sin\varphi,\hspace{1pt}\cos\varphi,0,0]^{\mathsf{T}},\\[2pt]
		\boldsymbol{e}^{\mathsf{(2)}}_A &= [0,0,0,\cos\varphi,\hspace{1pt}\sin\varphi,0,0]^{\mathsf{T}},
	\end{align} respectively.

\end{document}